\documentclass[a4paper,UKenglish,cleveref, autoref, thm-restate]{article}
\usepackage[a4paper, total={6in, 10in}]{geometry}
\usepackage{amsfonts}

\usepackage{todonotes}
\usepackage{algorithm}
\usepackage{algpseudocode}
\usepackage{multirow}
\usepackage{makecell}
\usepackage{hyperref} %
\usepackage{amsmath}
\usepackage{amsthm}
\usepackage{authblk}

\definecolor{teagreen}{rgb}{0.82, 0.94, 0.75}
\definecolor{azureblue}{rgb}{0.94, 1.0, 1.0}
\definecolor{flamingopink}{rgb}{1.0, 0.8, 0.8}
\definecolor{nadiacolor}{rgb}{0.9, 0.9, 1.0}
\newcommand{\cryptograph}{Crypto'Graph}

\newcommand{\V}{\mathcal{V}}
\newcommand{\E}{\mathcal{E}}

\newcommand{\Gstar}{G^\star}
\newcommand{\Gsquare}{G^2}
\newcommand{\candidates}{\textit{candidate}}
\newcommand{\noncandidates}{\textit{non\_candidate}}
\newcommand{\DegreeCompletionAttack}{\textsc{DegreeCompletionAttack}}
\newcommand{\NeighborCompletionAttack}{\textsc{NeighborCompletionAttack}}
\newcommand{\DegreeMatchingAttack}{\textsc{DegreeMatchingAttack}}
\newcommand{\NeighborMatchingAttack}{\textsc{NeighborMatchingAttack}}
\newcommand{\DegreeCombinationAttack}{\textsc{DegreeCombinationAttack}}
\newcommand{\TriangleAttack}{\textsc{TriangleAttack}}
\newcommand{\BicliqueAttack}{\textsc{BicliqueAttack}}

\newtheorem{problem}{Problem}
\newtheorem{proposition}{Proposition}
\newtheorem{theorem}{Theorem}

\newcommand{\norm}[1]{\left\lVert#1\right\rVert}

\begin{document}

\title{GRAND : Graph Reconstruction from potential partial Adjacency and Neighborhood Data}

\author[1]{Sofiane Azogagh}
\author[1]{Zelma Aubin Birba}
\author[2]{Josée Desharnais}
\author[1]{Sébastien Gambs}
\author[1]{Marc-Olivier Killijian}
\author[2]{Nadia Tawbi}
\affil[1]{Université du Québec À Montréal, \texttt{\{azogagh.sofiane, birba.zelma\_aubin\}@courrier.uqam.ca}, \texttt{\{gambs.sébastien, killijian.marc-olivier.2\}@uqam.ca}}
\affil[2]{Université Laval, \texttt{\{josee.desharnais, nadia.tawbi\}@ift.ulaval.ca}}

\date{}

\maketitle

\def\thefootnote{}
\footnotetext{This work is supported by the DEEL Project CRDPJ 537462-18 funded by the National Science and Engineering Research Council of Canada (NSERC) and the Consortium for Research and Innovation in Aerospace in Québec (CRIAQ), together with its industrial partners Thales Canada inc, Bell Textron Canada Limited, CAE inc and Bombardier inc. \url{https://deel.quebec}}

\def\thefootnote{\arabic{footnote}}

\begin{abstract}
Cryptographic approaches, such as secure multiparty computation, can be used to compute in a secure manner the function of a distributed graph without centralizing the data of each participant. 
However, the output of the protocol itself can leak sensitive information about the structure of the original graph.
In particular, in this work we propose an approach by which an adversary observing the result of a private protocol for the computation of the number of common neighbors between all pairs of vertices, can reconstruct the adjacency matrix of the graph. In fact, this can only be done up to co-squareness, a notion we introduce, as two different graphs can have the same matrix of common neighbors. We consider two models of adversary, one who observes the common neighbors matrix only, and a knowledgeable one, that has a partial knowledge of the original graph.
Our results demonstrate that secure multiparty protocols are not enough for privacy protection, especially in the context of highly structured data such as graphs. The reconstruction that we propose is interesting in itself from the point of view of graph theory. 
\end{abstract}

\maketitle

\section{Introduction}

Graphs have emerged as the predominant format for representing relational data, as they naturally capture both the relationships and structures inherent in such datasets.  
Indeed, from social networks~\cite{Wilson09} to biological systems~\cite{pavlopoulos2011using}, the interconnection of entities can be easily visualized and understood through graphs. 
However, as data becomes increasingly distributed, new challenges arise with respect to its analysis. 
For example, in a scenario in which the structure of a graph is distributed across multiple parties (\emph{e.g.}, a social network distributed across multiple servers or a shared knowledge graph), the objective might be to study this structure without centralizing this data and without each party disclosing the private details of their segment. 

For instance, in this setting, edge prediction and edge removal techniques can be used to improve the quality of local knowledge. 
More precisely, edge prediction aims at inferring the existence of an edge between two vertices, with typical use cases being in recommendation systems or in social networks~\cite{hasan2011survey} but also in anomaly detection, influence analysis and community detection~\cite{daud2020applications}. 
In contrast, edge removal has the objective of deciding if an edge between two vertices should not exist. 
This can be used, for example, as a counter-measure to data poisoning,  in which adversarial edges may have been introduced by an adversary to compromise the graph integrity~\cite{Zugner18, Wu19, Xu23}.
One of the classical techniques for predicting the existence/non-existence of an edge is based on the computation of the number of common neighbors and infers that two vertices sharing several, respectively few, neighbors should probably be linked, respectively unlinked. 

\begin{figure}
    \centering
    \includegraphics[width=0.5\linewidth]{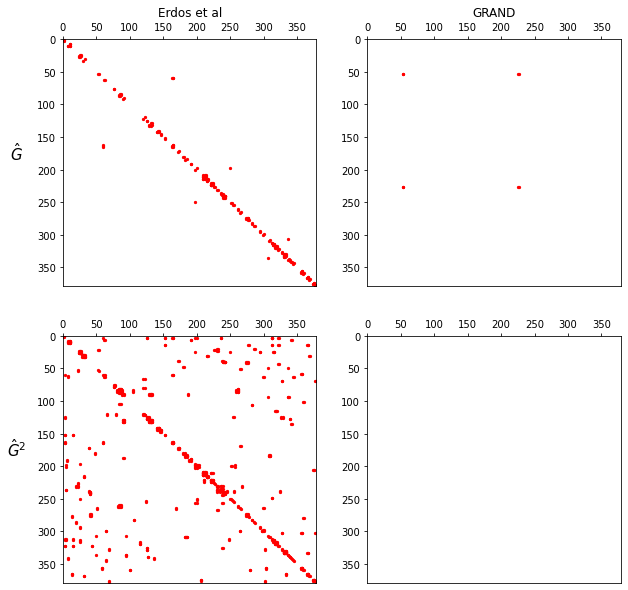}
    \caption{Reconstruction errors on Netscience, in terms of the adjacency matrix and the common neighbors matrix. $\hat{G}$ denotes the adjacency matrix of the reconstructed graph, and $\hat{G}^2$ denotes its square. Red dots represent mismatches with regard to the original matrices.}
    \label{fig:enter-label}
\end{figure}

For instance, \cryptograph{} \cite{CryptoGraph} enables the oblivious computation of the number of common neighbors between a pair of vertices from a graph distributed among two participants. 
From this number of common neighbors, each participant can decide whether the two vertices should share an edge in their own local graph, and thus perform local edge prediction or removal. 
The usage of such a private protocol to compute the common neighbors matrix on a distributed graph can be motivated by its applications in the context of social media platforms trying to improve their recommendation algorithms by computing the number of common friends of their users across their joint networks, as explored previously in the literature~\cite{Demirag23, CryptoGraph}.

While in such protocols, cryptography can be used to compute the number of common neighbors between each pair of vertices, an honest-but-curious participant can try to infer some knowledge about the graph of the other participant. 
In this work, we precisely aim at characterizing what can be learned by an honest-but-curious participant from the output of the protocol as well as their knowledge of their own local graph. We also explore the setting in which the adversary only has access to the output of the computation of the number of common neighbors between vertices, and tries to infer the structure of the graph.

Imagine a scenario where two pharmaceutical laboratories seek to collaboratively predict previously undetected or unobserved drug-drug interactions~\cite{zhang2017predicting,vilar2013detection}, while safeguarding the privacy of their proprietary data. Each laboratory possesses a list of medications and a graph of drugs where a drug is linked to another if its effectiveness or toxicity is influenced by the co-administration of the other. By the end of the protocol, for each pair of medication $(X, Y)$ on the merged medication graphs (also known as the interaction profile fingerprint~\cite{vilar2013detection}), both laboratories obtain the number of drugs that influence both $X$ and $Y$. This scenario highlights the potential for one laboratory to exploit the output of the protocol to infer proprietary data from the other, identifying medications that interact favorably in the other's graph. Such instances emphasize the necessity for robust privacy measures in the context of multiparty computation protocols, particularly within sensitive domains like pharmaceutical research.

\paragraph{Contributions.}

In this paper, we develop a graph data reconstruction attack based on the number of common neighbors of its vertices. To achieve this, we design topological deterministic attacks that infer some information on the graph structure based on the number of common neighbors. The knowledge of the graph obtained from these topological attacks is then used to enhance the spectrum-based attack introduced by Erd\H{o}s et al~\cite{Erdös12}. This study led us to introduce a new notion of equivalence between graphs based on their common neighbors matrices, as well as an appropriate metric to fairly compare the reconstructed graph and the original one. By considering two types of adversaries, a knowledgeable and an unknowledgeable one, we demonstrate through experiments on real-world datasets that it is possible to perfectly reconstruct certain graphs even without a prior knowledge on the target graph. We also show how a partial knowledge of the structure of the graph can be leveraged to obtain an even better reconstruction on other instances.

\paragraph{Outline.} We begin this paper by formalizing the problem we are tackling in Section~\ref{sec:problem-statement}. Then we review some work in the literature related to the problem  in Section~\ref{sec:related-work}. After that, in Section~\ref{sec:theoretical-building-blocks}, we present the theoretical background needed to understand our attacks. Specifically, we present some properties inferred from the common neighbors matrix that we use to reconstruct the graph. We then present the algorithms composing our attack named \texttt{GRAND} in Section~\ref{sec:grand} and the experimental results in Section~\ref{sec:experiments}. Finally, we conclude the paper in Section~\ref{sec:conclusion} with a discussion on future work.

\section{Problem definition}
\label{sec:problem-statement}

\paragraph{System overview.}

In this work, we are considering an undirected graph $G = (\V,\E)$ with no self-loop.
The (open) neighborhood of a vertex $v \in \V$ on $G$ is denoted as $\Gamma(v)$ and includes all vertices that share an edge with $v$ on $G$. 
For any two vertices $u$ and $v$ in $\V$, the  \textit{common neighborhood} of $u,v \in V$ on $G$ is the set $\Gamma(u) \cap \Gamma(v)$. 
In particular, we are interested in the \emph{common neighbor matrix},
in which each entry $(u,v)$ contains the number of common neighbors of $u$ and $v$ in the graph $G$. 
It is easy to see that this matrix is in fact the result of the matrix multiplication of the adjacency matrix of $G$ by itself, which we denote by $G^2$.
Table~\ref{table-notation} summarizes the notations that we will be using in the paper.

\begin{table}[h!]
\centering
\begin{tabular}{|l|l|}
\hline
$\V$ & set of vertices (\emph{i.e.,} vertices) \\ \hline
$\E$ & set of edges (\emph{i.e.,} links)  \\ \hline
$G=(\V,\E)$ & graph of vertices $\V$ and edges $\E$   \\ \hline
$\Gamma(x) \subseteq\V$ & \makecell[l]{neighbors of $x$ in $G$ (\emph{i.e.}, $\{v \in\V\mid (v,x)\in\E$)}  
\\ \hline
$\Gamma^\star(x) \subseteq\V$ & neighbors of $x$ in $\Gstar$   
\\ \hline
$G^2$ & \makecell[l]{common neighbors matrix of $G$ \\ ( \emph{i.e.}, $G^2(u,v) = |\Gamma(u) \cap \Gamma(v)|$))} \\ \hline
$\Gamma^2(x)$ & \makecell[l]{neighbors of $x$ in $G^2$, that is, vertices that \\have at least one common neighbor  with $u$} \\ \hline
$M\vert_{E}$ & matrix $M$ restricted to $E$ \\ \hline
$\rho$ & \makecell[l]{Proportion of existing and non-existing \\ edges known by attacker} \\ \hline 
\end{tabular}
\caption{Summary of the notations used in this paper.}
\label{table-notation}
\end{table}

\paragraph{Adversary models.}
We consider two types of adversaries. 
The first one corresponds to an external attacker obtaining the result of a private computation of the common neighbors matrix on the 
graph $G$. 
This type of attacker does not have any prior knowledge about $G$ and aims at reconstructing it solely based on $G^2$. 
This adversary model has already been explored in works such as \cite{Erdös12}. We designate this type of attacker as \emph{unknowledgeable}.

The second adversary model is motivated by the more general setting in which  the attacker has some prior knowledge on $G$ (\emph{e.g.}, the existence or non-existence of some of the edges of the graph $G$). For example, this is the case for a participant to the private computation of the common neighbors matrix on the distributed graph $G$. 
Such an adversary could be, for instance, an honest-but-curious participant who inputs their subgraph of the graph $G$ to the private protocol to obtain $G^2$ as output, and tries to reconstruct $G$ from the knowledge of their subgraph and $G^2$. 
We express the prior knowledge of the attacker as a set of edges that exist in the graph denoted $\E_1$, as well as a set of edges that do not exist called $\E_0$. The rest of the possible connections between vertices are considered  unknown.
It is easy to notice that when $\E_1 = \E_0 = \emptyset$, the second adversary model corresponds to the first, hence leading to a more general problem.
We call such an attacker a \emph{knowledgeable} attacker.

\paragraph{Problem statement.}
From the above adversary models, we devise the following problem statement :
\begin{problem}[Reconstructing $G$ from $G^2, \E_0, \E_1$]
\label{problem}

Let $G$ be a graph, reconstruct $G$ from the knowledge of its common neighbors matrix $\Gsquare$ and the lists $\E_0$ and $\E_1$ of non-existing and existing edges in $G$.
\end{problem}

Some work in the literature presented in the following section, tackle the specific case of reconstruction of $G$ by an unknowledgeable attacker.

We suspect that this problem may be NP-hard, as it is really close to known NP-complete problems, for example the Intersection Pattern problem~\cite{CHVETAL1980249} and the Square Root Graph problem ~\cite{motwani1994computing} (in the latter, the square root of a graph is different from the definition we state in this paper, as is discussed in the next section). We have not found a theoretical analysis of this exact problem, which is not standard in graph theory, as isomorphic graphs do not have the same square matrix in general. We will discuss this matter in subsection~\ref{sec:non-unicity}.

\section{Related Work}
\label{sec:related-work}

The reconstruction of graphs has received much attention over the years, with the motivation coming from different use cases, such as recommendation in social networks~\cite{Wilson09,Nitai10}, uncovering hidden interactions between proteins or molecules~\cite{Muzio20} as well as the discovery of unknown relationships between criminal organizations~\cite{McAndrew21}. 
While the objective remains to reconstruct the network as accurately as possible, the various approaches differ in their initial knowledge and the adopted methodologies.

However, the problem of recovering $G$ from $G^2$ as defined in Section~\ref{sec:problem-statement} has received much less attention.  %
In the field of linear algebra, the  problem of finding the square root of a matrix (the adjacency matrix of $G$ is one square root of the matrix $G^2$) is well-studied problem.
However, the matrix studied does not necessarily correspond to a binary matrix (\emph{i.e.}, the adjacency matrix of a graph).  
By construction,  $G^2$ is  a positive semi-definite matrix (psd), because there exists a matrix such that $BB^T = G^2$, this matrix being $B:=G=B^T$. 
As $G$ is symmetric, a theorem (see~\cite{horn2012matrix} Theorem 7.26) states that there exists a unique symmetric positive semi-definite matrix that serves as the {square root} of this $G^2$.
However this theorem does not solve our problem because $G$, the square root that we are looking for, may not be itself psd.
Furthermore, the psd solution is not even a binary matrix in general.

In the field of graph theory, all graph theoretical references define $G^2$ in a different manner. 
Many define it as binary matrix. For instance,  in~\cite{MOTWANI199481} the entries of the matrix represent the existence of a path of length at most 2 between the vertices, which is $G\cup G^2$ with 0-1 entries, while in~\cite{bai2024characterizing} it represent the existence of a path of length exactly 2 between the vertices.
As a result, the problem considered is quite different. In our setting, the value is the number of paths of length exactly 2. So $G^2$ is not always a binary matrix and $G$ is not a subgraph of it.

To the best of our knowledge, the work closest to ours is~\cite{Erdös12}, which proposes a method for the reconstruction of the adjacency matrix of a (bipartite) graph, knowing the common neighbors information between all pairs of vertices. 
Their approach, which we revisit and improve in our work,  relies on the spectral domain of the adjacency matrix. Indeed,  it exploits the properties of the singular value decomposition of the common neighbors matrix to iteratively reconstruct an approximation of the desired graph. 
However, relying on the graph's spectral domain makes them susceptible to reconstruct a co-spectral graph. 
Co-spectral graphs~\cite{VANDAM2003241} exhibit identical spectra but are non isomorphic (by definition), thereby making it impossible to uniquely reconstruct the original graph based solely on its eigenvalues.

Remark that the diagonal of $G^2$ contains the degrees of the vertices of $G$.
The problem of reconstructing $G$ from its sequence of degrees 
has received a lot attention~\cite{zoltan12, cloteaux16} but the proposed algorithms to solve this problem potentially have to explore an exponential number of solutions~\cite{zoltan12}. 
This problem is more general than the one we solved in this paper, since we have more information.
Also, in some settings, the diagonal might not be available to the attacker, although we do not study this scenario.

This paper goes beyond existing works by considering a more powerful adversary than previously studied in works such as~\cite{Erdös12, cloteaux16}.
We propose a solution that achieves the best known reconstruction performance of an undirected graph solely based on its common neighbors matrix.
In addition to that setting, we study the adversarial model in which  besides the common neighbors matrix, the adversary possesses some partial knowledge of the original graph. 
Our experiments demonstrate that this additional knowledge grants the adversary a higher capacity to reconstruct the graph.

\section{Theoretical building blocks}
\label{sec:theoretical-building-blocks}

\begin{figure}[h!]
    \centering
    \includegraphics[width=0.7\linewidth]{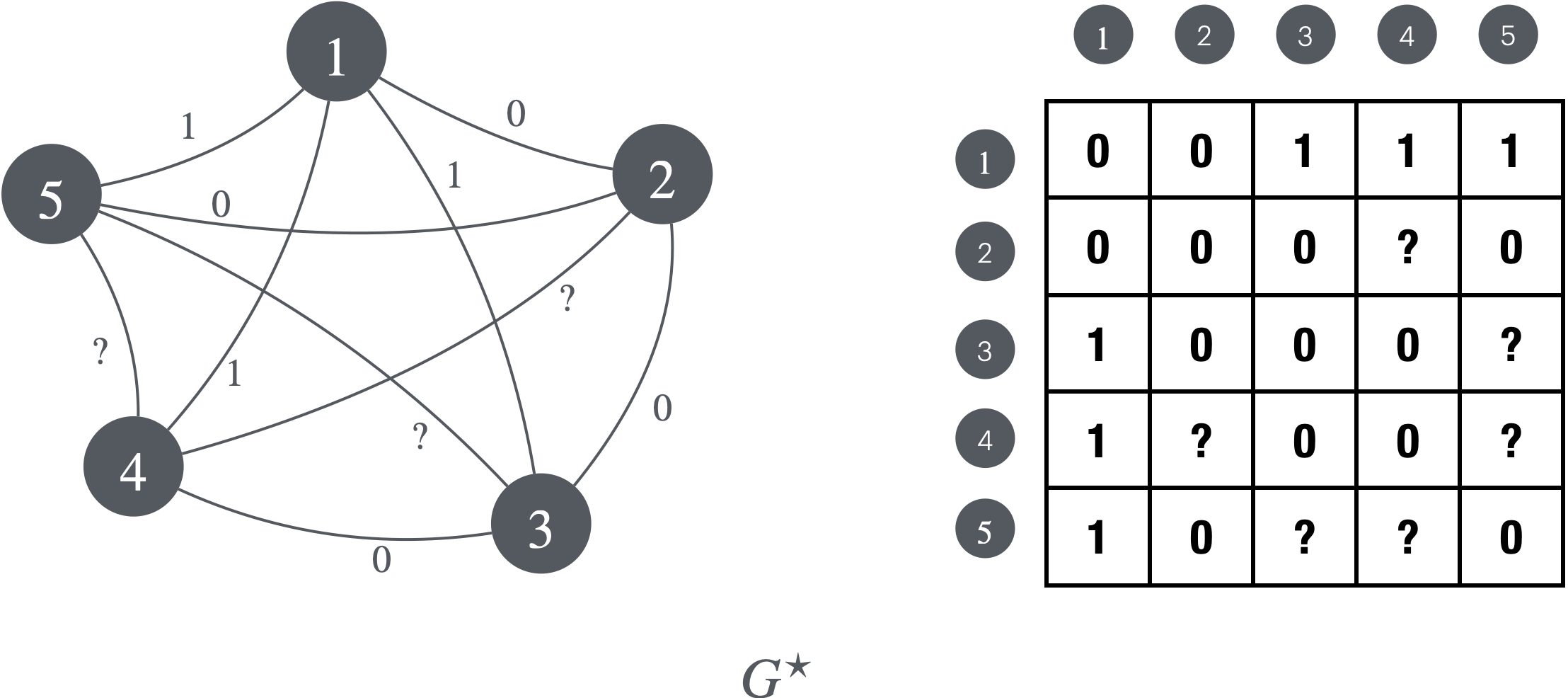}
    \caption{Example of a graph $\Gstar$ and its adjacency matrix.}
    \label{fig:example-gstar}
\end{figure}

Recall that our objective is, given some partial knowledge of the adjacency on $G$, 
as well $G^2$ the common neighbors matrix, to reconstruct $G$.
Our attack will reconstruct $G$ in different steps with the reconstructed graph being referred as $\Gstar$. Figure \ref{fig:example-gstar} depicts our representation of the reconstructed graph $\Gstar$ during the attacks. It can be seen as a complete graph where the edges are labeled as 0, 1 or ? respectively if in $G$, they exist, do not exist, or we do not know.

During the reconstruction, the cells of $\Gstar$ are updated to reflect the information learned. 
Hence, cells that have  value  $?$s are turned into $1s$ or $0s$ depending on the inference performed in $\Gstar$ when we learn the existence or absence of an edge in $G$.
The final reconstructed graph obtained at the end of the attacks, which we denote by $\hat{G}$, contains only 0s and 1s.

\subsection{Domains of study}

Consider the adjacency matrix of the graph $G$, denoted as $M_G$. It can be expressed as: 
\begin{equation}
    \label{eq:eigen-decomp}
    M_G = U \Lambda U^T
\end{equation}
where $U$ is an orthogonal matrix and $\Lambda$ is a diagonal matrix containing the eigenvalues of $M_G$. This decomposition highlights that there are two equivalent representations of the graph information: the left-hand side of Equation~\eqref{eq:eigen-decomp}, which we call the \emph{topology domain}, and the right-hand side, which we call the \emph{spectral domain}. The equation demonstrates that these two domains encode the same information in different forms. As mentioned in~\cite{van2023graph}, similar to how signals can be analyzed in both time and frequency domains via Fourier transforms, graphs can be studied in both topological and spectral domains, with each domain better suited for analyzing different properties of the graph.

The topology domain is the most intuitive representation of a graph, where vertices and their connections can be directly visualized and understood by humans. In this domain, we can easily reason about structural properties like paths, cycles, and connectivity patterns between vertices. This representation naturally lends itself to analyzing local properties and deriving insights about the graph's structure through direct observation.

The spectral domain, on the other hand, studies graphs through the eigenvalues and eigenvectors of their adjacency matrices. While less intuitive for human understanding, this domain has given rise to spectral graph theory - a rich field that connects graph properties to linear algebra. The interpretation of eigenvalues and eigenvectors is not immediately obvious, but their study has revealed deep connections to many graph properties like connectivity, clustering, and symmetry.

\subsection{Non-unicity of the solution}
\label{sec:non-unicity}

\paragraph{Spectral.}
As we will detail in subsection~\ref{subsec:singular_val_decomp}, $G^2$'s spectrum   is linked to $G$'s spectrum, as the former's eigenvalues are the squares of the latter's. This gives a strategy to find  $G$ through its eigenvalues that we exploit, in the same manner as  Erd\H{o}s et al.~\cite{Erdös12} do. However, even with the knowledge of the exact spectrum of a graph (which is not what we have here), one can end up with a graph that is co-spectral to the one they are looking for (and non isomorphic, by definition). Indeed, the spectral non-uniqueness has been demonstrated in previous research~\cite{VANDAM2003241,VANMIEGHEM202434}, as there exist multiple non-isomorphic graphs sharing the same eigenvalues or the same eigenvectors.

Interestingly, a different notion of  equivalence is at play here, that we call \emph{co-square equivalence}. The name of vertices, that is, their position in the matrix, is important when comparing the matrices of common neighbors. Hence, isomorphic graphs, in the classical sense, do not necessarily have the same matrix of common neighbors. We now elaborate on this.

\begin{figure}[h]
    \centering
    \includegraphics[width=0.7\linewidth]{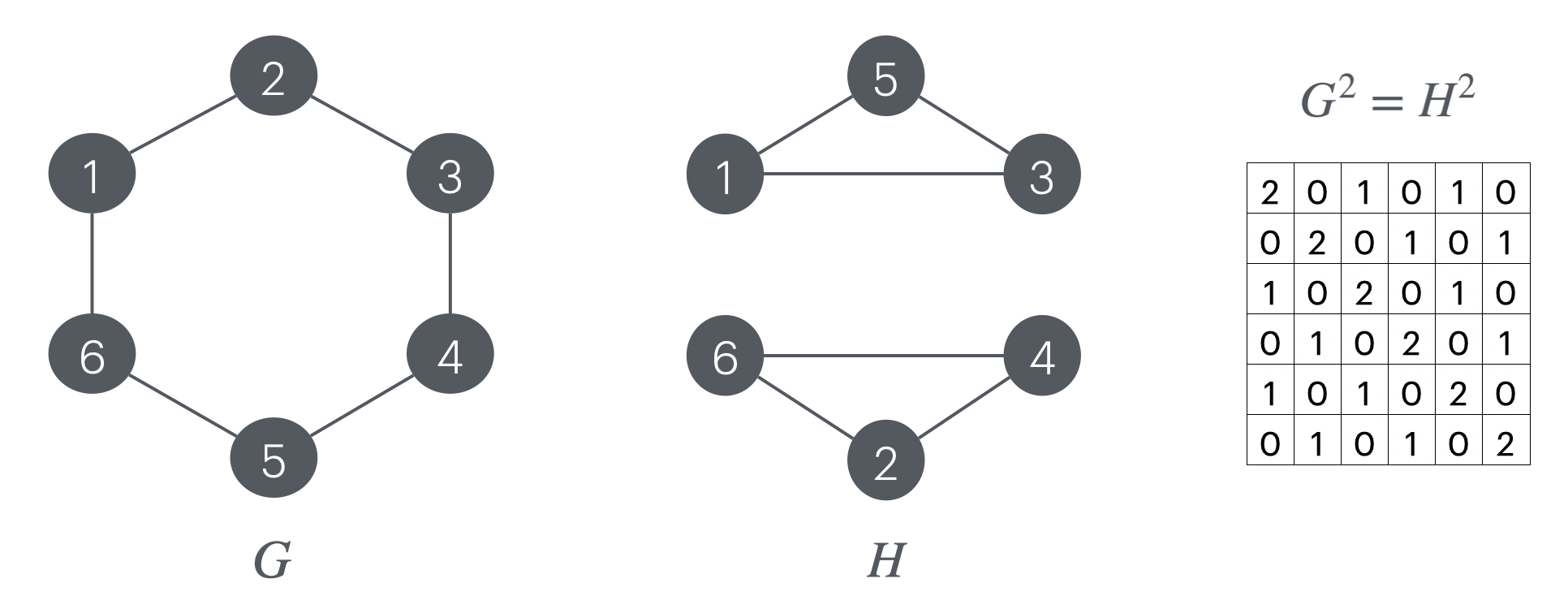}
    \caption{%
    Two  co-square (and non co-spectral) graphs}
    \label{fig:example-non-unicity}
\end{figure}

\paragraph{Topological.} 
In Figure~\ref{fig:example-non-unicity}, there are two graphs with the same common neighbor matrix, but   non isomorphic and  not co-spectral: the eigenvalues of $G$ and $H$ are respectively $\{ -2, -1, -1, 1, 1, 2\}$ and $\{-1, -1, -1, -1, 2, 2\}$. 
We call two graphs \emph{co-square} if they have the same common neighbors matrix without permuting any line or column. This definition is incomparable with isomorphism between graphs, as there are non-isomorphic graphs (Fig.~\ref{fig:example-non-unicity}) with the same matrix of common neighbors; conversely, there are isomorphic graphs (take a well chosen permutation of vertices of a graph with vertices of different degrees), 
that do not have the same matrix of common neighbors (but that are equal up to a symmetric permutation of lines and columns). %

The notion of co-squareness is also incomparable with  co-spectrality, as there are graphs that are not co-spectral and yet still share the same common neighbors matrix (take graphs with the same absolute value of eigenvalues). Conversely, there are co-spectral graphs that are not co-square: take co-spectral graphs with, again, vertices of different degrees and permute the position of these vertices in the corresponding matrix.  To the best of our knowledge, co-square graphs do not belong to any known category in graph theory. %

In summary, we are in presence of three notions of equivalences between graphs. Isomorphism and co-spectrality are incomparable by definition, and  co-squareness is also incomparable to both notions.

As non unicity can lure us to construct a co-square graph, a prior knowledge on the target graph's structure might help to distinguish which is the right one. 
For instance, the \emph{a priori} information that $(1, 3)$ does not exist in $G$ removes the possibility that $H$ would be the target graph. In our work,  such information is exploited by the knowledgeable adversary.

\subsection{Properties inference of $G$}
\label{sec:propositions}

Hereafter, we leverage the common neighbors information $G^2$, as well as the partial knowledge $\E_0, \E_1$ to deduce properties about $G$.
Recall that the partial reconstruction of $G$ is denoted by $\Gstar$ while the number of common neighbors between vertices $u, v \in \V$ is given by $G^2(u, v)$ and the degree of vertex $u$ is $G^2(u, u)$.

\subsubsection{Topological properties}

\begin{proposition}[Rows of $G^2$ and degrees of neighbors]
\label{prop:1}
    Let $u \in \V$ be a vertex on $G$. 
    Then, the sum of the degrees of its neighbors is equal to the sum of the $u^{th}$ row in $G^2$.
    \begin{equation}
        \forall u \in \V, \sum_{v \in \V} G^2(u,v) = \sum_{w \in \Gamma(u)} |\Gamma(w)|
    \end{equation}

\end{proposition}

This is not a complete characterization of a common neighbors matrix, as there are matrices satisfying this property that do not correspond to any graph. 
However, we can exploit Proposition~\ref{prop:1} to infer some existing and non-existing edges in $G$ that are not known in $\Gstar$, 
which is the objective of Algorithm~\ref{alg:degree-combination}.

\begin{proposition}[Completeness of a neighborhood in $\Gstar$]
\label{prop:2}
    Let $u \in \V$ be a vertex of $G$. 
    If $u$ has the same degree on $G$ and $\Gstar$, then the neighborhood of $u$ on $\Gstar$ is complete.
    \begin{equation}
        \forall u \in \V, |\Gamma^{\star}(u)| =  |\Gamma(u)| \implies \forall v \in \V \backslash \Gamma^{\star}(u), (u, v) \notin \E 
    \end{equation}
    Similarly, if two vertices $u$ and $v$ have the same number of common neighbors on $G$ and $\Gstar$, then the common neighborhood of $u$ and $v$ is complete.
    \begin{equation}
        \begin{split}
            \forall u, v \in \V, |\Gamma^{\star}(u) \cap \Gamma^{\star}(v)| = |\Gamma(u) \cap \Gamma(v)| \implies  \\ [ \forall w \in \Gamma^{\star}(u) \backslash \Gamma^{\star}(v),
                (w, v) \notin \E]
        \end{split}
    \end{equation}
\end{proposition}

We exploit the information brought by these propositions in Algorithms ~\ref{alg:degree-matching} and \ref{alg:neighbor-matching}.

\begin{proposition}[Completing a neighborhood in $\Gstar$]
\label{prop:3}
    Let $u \in \V$ be a vertex on $G$. If the following conditions are met :
    \begin{enumerate}
        \item $u$ is missing $k$ neighbors (\emph{i.e.}, $|\Gamma(u)|-|\Gamma^{\star}(u)| = k$);
        \item there are exactly $k$ vertices $v_1, \dots, v_k$ such that we do not know if $(u, v_i)$ exists.
    \end{enumerate}
    Then all edges $(u, v_i)$ must exist in $G$.

    Similarly, for common neighborhoods, if:
    \begin{enumerate}
        \item $u$ and $v$ are missing $k$ vertices in their common neighborhood ($ \emph{i.e.},  |\Gamma(u) \cap \Gamma(v)|-|\Gamma^{\star}(u) \cap \Gamma^{\star}(v)| = k$);
        \item there are exactly $k$ vertices $w_1, \dots, w_k$ such that we do not know if $(u, w_i)$ exists.
    \end{enumerate}
    
    Then all edges $(u, w_i), (v, w_i)$ must exist.
\end{proposition}

Proposition ~\ref{prop:3} can be seen as the complementary of Proposition \ref{prop:2}, in the sense that the latter identifies some $0$s in $\Gstar$ based on the number of missing neighbors (or common neighbors), while the former uses the same information to identify some $1$s. 
Proposition~\ref{prop:3} is directly used in Algorithms ~\ref{alg:degree-completion} and \ref{alg:neighbor-completion}. 

\begin{proposition}[Triangles]
\label{prop:triangle}
    Let $u,v$ be two vertices on $\V$, if the following conditions are satisfied
    \begin{enumerate}
        \item $u$ and $v$ are connected in $G^{\star}$ (\emph{i.e.}, $(u,v) \in \E^{\star}_1$)
        \item $u$ and $v$ have at least one common neighbor (\emph{i.e.}, $G^2(u,v) >0$)
    \end{enumerate}
    then $u,v$ are in $k=G^2(u,v)$ triangles.
    Their common neighbors are then  vertices $w_1,\dots,w_k \in \V$ such that 
$$\forall i \in \{1,\dots,k\}, G^2(u,w_i)>0 \text{ and } G^2(v,w_i)>0,$$
\end{proposition}

 Algorithm~\ref{alg:triangle} implements the findings of Proposition~\ref{prop:triangle}.

\begin{proposition}[Bi-cliques\footnote{A bi-clique, or complete bipartite graph, is a bipartite graph where each vertex is connected to all the vertices in the other partition.}]
\label{prop:rectangle}
    Let $ U=\{u_1, \dots, u_p\}$ and $ V=\{v_1, \dots, v_q\}$ be two sets of vertices in $\V$. If the following conditions are satisfied 
    \begin{enumerate}
        \item The neighborhood of $u_1$ is complete : $\Gamma(u_1) = \Gamma^{\star}(u_1) = \{v_1, \dots, v_q\}$ 
        \item Each vertex $u_i$, $i \in \{2, \dots, p\}$ shares exactly $q$ common neighbors with $u_1$.
    \end{enumerate}
    then $U \cup V$ forms a bi-clique, in which $U$ and $V$ are the associated partitions. 
\end{proposition}

Algorithm~\ref{alg:bi-clique} searches vertices with completely known neighborhoods in $G^{\star}$ in order to identify bi-cliques using this proposition.

The following proposition is interesting in itself, but does not lead to an algorithm that we have implemented.
\begin{proposition}[Bipartition or disconnectedness of G ]
    \label{prop:disconnectedness}
    $\Gsquare$ is disconnected if, and only if, $G$ is either  disconnected or bipartite.
\end{proposition}

An example of such a case is given in Figure \ref{fig:example-non-unicity}, where both a bipartite graph $G$ and a disconnected graph $H$ have the same common (disconnected) neighbors matrix. This example also explain why we have not implemented any corresponding algorithm. Given a disconnected matrix $G^2$ (which can be determined in linear time), one cannot know for sure if the wanted graph is disconnected or if it is bipartite. There are  examples where only a bipartite graph exists, so one cannot focus on each of the components to find a co-square graph (which is of course the best one can do if no prior knowledge is given).

\subsubsection{Spectral properties}
\label{subsec:singular_val_decomp}

We start by presenting the Singular Value Decomposition (SVD) that we mentioned before for (symmetric) matrices, elaborate on how one can gain information of the singular values of $G$ from $G^2$. Note that as $G^2$ is a real, symmetric matrix, its SVD coincides with its eigendecomposition (i.e the singular values are the eigenvalues). We can easily express $G^2$ as :
$$
G^2 = U \Lambda U^T
$$
where $\Lambda$ is a diagonal matrix containing the eigenvalues $\lambda_1, \dots, \lambda_n$ and $U$ is an orthogonal matrix containing the eigenvectors $u_1, \dots, u_n$ of $G^2$. As $U$ is orthogonal we have $U^T = U^{-1}$, which trivially implies that:
$$
G = U \Sigma U^T
$$
where $\Sigma$ is a diagonal matrix containing the eigenvalues of $G$ (i.e., $\sigma_1, \dots, \sigma_n$) which can be either the positive or negative square roots of the eigenvalues of $G^2$ (i.e., $\pm\sqrt{\lambda_1}, \dots, \pm\sqrt{\lambda_n}$). So the problem of reconstructing an undirected graph $G$ from the spectrum of its common neighbors matrix $G^2$ can be solved by finding the right sign for each eigenvalue. This sign assignment problem is likely what could make the reconstruction problem NP-hard, since there are $2^n$ possible sign combinations for the eigenvalues. Indeed, for each eigenvalue $\lambda_i$ of $G^2$, we need to choose between $+\sqrt{\lambda_i}$ and $-\sqrt{\lambda_i}$ as the corresponding eigenvalue of $G$, leading to an exponential number of possibilities to explore. Erd\H{o}s et al, in \cite{Erdös12} provide an heuristic method exploiting this insight. Their \textsc{Greedy} algorithm chooses the right eigenvalue based on the following results from Eckart and Young theorem~\cite{eckart1936approximation} :

\begin{theorem}[Eckart and Young]
    The best rank-$k$ approximation of $G$ in the Frobenius norm $\norm{\cdot}_F$ is given by 
    $$\sum_{i=1}^k u_i \sigma_i u_i^T$$
    where $\sigma_i$ is the $i$-th largest eigenvalue of $G$ and $u_i$ is the corresponding eigenvector.
\end{theorem}

This powerful theorem allows to reconstruct $G$ iteratively by choosing the sign of $\sqrt{\lambda_i}$ that minimizes the Frobenius distance at each step for $i=1,\dots, n$. Moreover, the early iterations are the most important ones as the largest eigenvalues (which contain the most information) are the first ones. Therefore, any prior knowledge of the graph should be used iteratively to "guide" the reconstruction in the right direction, and the more knowledge we have, the better we can guide the reconstruction. This is the idea exploited in Algorithm~\ref{alg:SpectralAttack}.

\section{GRAND}
\label{sec:grand}

In the following sections, we propose several algorithms for the reconstruction of $G$ given its common neighbors matrix $G^2$. 
We also describe how this reconstruction can be enhanced from the knowledge of the existing and non-existing edges $\E_1, \E_0$. 
More precisely, we first describe deterministic algorithms that can infer \emph{exactly} the existence or non-existence of edges in $G$ based on topological properties. 
In some cases, there remains unknown relations between vertices.
To further improve the reconstruction, we then present heuristics for the reconstruction of $G$ based on the spectrum of $G^2$ while leveraging the partial reconstruction obtained after the topological attacks. 
\subsection{Topological attacks}

We now present algorithms that leverage the propositions described previously to reconstruct $G$. 
As discussed before, these algorithms recover $G$ in an exact manner w.r.t.~$G^2$, that is, they make sure that the square of the constructed graph is equal to $G^2$.
Some of them use the prior knowledge ($\E_0, \E_1$). 
We start by incorporating the prior knowledge (if any) into the reconstructed graph $\Gstar$ as follows:
\begin{itemize}
    \item All existing edges in the prior knowledge get a label of 1 in $\Gstar$. 
    That is, $\forall (u, v) \in \E_1, \Gstar(u, v) \gets 1$.
    \item Similarly, non-existing edges in the prior knowledge get assigned 0 in $\Gstar$. $\forall (u, v) \in \E_0,\Gstar(u, v) \gets 0$.
    \item $G$ does not contain self-loops, so $\forall u \in V, \Gstar(u, u) \gets 0$.
    \item All the other possible edges get a label of $?$.
\end{itemize}

\begin{figure}
    \centering
    \includegraphics[width=0.7\linewidth]{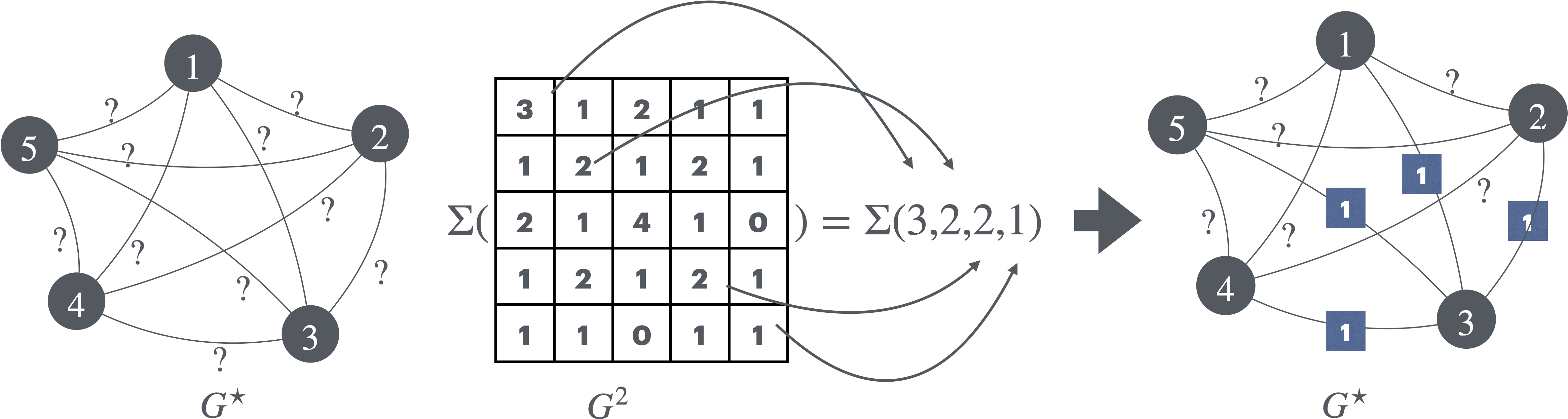}
    \caption{\DegreeCombinationAttack. The sum of the row of vertex 3 in $\Gsquare$ is equal to the sum of the degrees of 1, 2, 4 and 5. Thus, the edges (1, 3), (2, 3), (4, 3), (5, 3) do exist in $G$.}
    \label{fig:degree-combination}
\end{figure}

\paragraph{\DegreeCombinationAttack}
Proposition \ref{prop:1} allows the following inference : given a vertex $u$, if there is only one candidate set of vertices in $\V$, $S = \{w_1, \dots, w_{|\Gamma(u)|}\}$ such that $\sum_{v \in \V} G^2(u,v) = \sum_{w_i \in S} G^2(w_i, w_i)$, then $(u, w_i) \in G$. 
One can use this observation to identify potential vertices that belong to the neighborhood of a certain vertex, as depicted in Algorithm ~\ref{alg:degree-combination}, lines 8-20. Similarly, we can also use this proposition to infer non existing edges in $G$, as done in lines 7 and 21-23: given two vertices $u, v$,  if the degree of a vertex $v$ exceeds the sum of the $u^{th}$ row in $\Gsquare$ (that is, without the degree of $u$), then $(u, v) \notin \E$. Figure ~\ref{fig:degree-combination} gives an illustration of this attack by infering the presence of edges in $G$.

\begin{algorithm}[h!]
    \begin{algorithmic}[1]
        \caption{Determining existence of edges of $G$ from the degrees of vertices.}
        \Require Partial reconstruction $\Gstar$, common neighbors matrix $G^2$
        \label{alg:degree-combination}
        \Function{\DegreeCombinationAttack}{$\Gstar, G^2$}
            \For{$u \in \V$}
                \State $s \gets \Sigma_{v \in \V }G^2(u,v)$ \Comment{Sum of u-th row}
                \State $d \gets G^2(u, u)$ \Comment{Degree of vertex u}
                \State $S = \{\}$ \Comment{Candidate set}
                \State $\candidates = \{v\mid \Gsquare(v, v) < s - d\}$

                \State $\noncandidates = \V \backslash \candidates$
                 \ForAll{$C \subseteq \candidates$ of size $d$} \Comment{Search  set}%
                    \If{$S = \{\}$}
                        \If{$\Sigma_{w \in C}G^2(w, w) = s$} 
                            \State $S \gets C$ \Comment{Candidate set found}
                        \EndIf
                    \Else \Comment{Multiple sets found}
                        \State $S \gets \{\}$
                        \State \textbf{break}
                    \EndIf
                 \EndFor

                 \ForAll{$v \in S$} \Comment{Update $\Gstar$}
                    \State $\Gstar(u, v) \gets 1$
                 \EndFor

                 \ForAll{$v \in \noncandidates$}
                    \State $\Gstar(u, v) \gets 0$
                 \EndFor
            \EndFor
        \State \Return $\Gstar$
        \EndFunction
    \end{algorithmic}
\end{algorithm}

Algorithm ~\ref{alg:degree-combination} iterates over all subsets of size the degree of the target vertex $u$, which might lead to a very high complexity. For optimization purposes, we run this algorithm only on vertices with degree at most 2. 

For this attack to work, one needs to have the degree of each vertex in $G$. When $G^2$ is fully available to the attacker, this is given by the diagonal of $G^2$ ( $G^2(u, u) = |\Gamma(u)|$ ). Moreover, this attack does not assume that the attacker has prior knowledge on $G$, allowing them to infer information about $G$ even when $\E_0 = \E_1 = \emptyset$.

\begin{figure}
    \centering
    \includegraphics[width=0.7\linewidth]{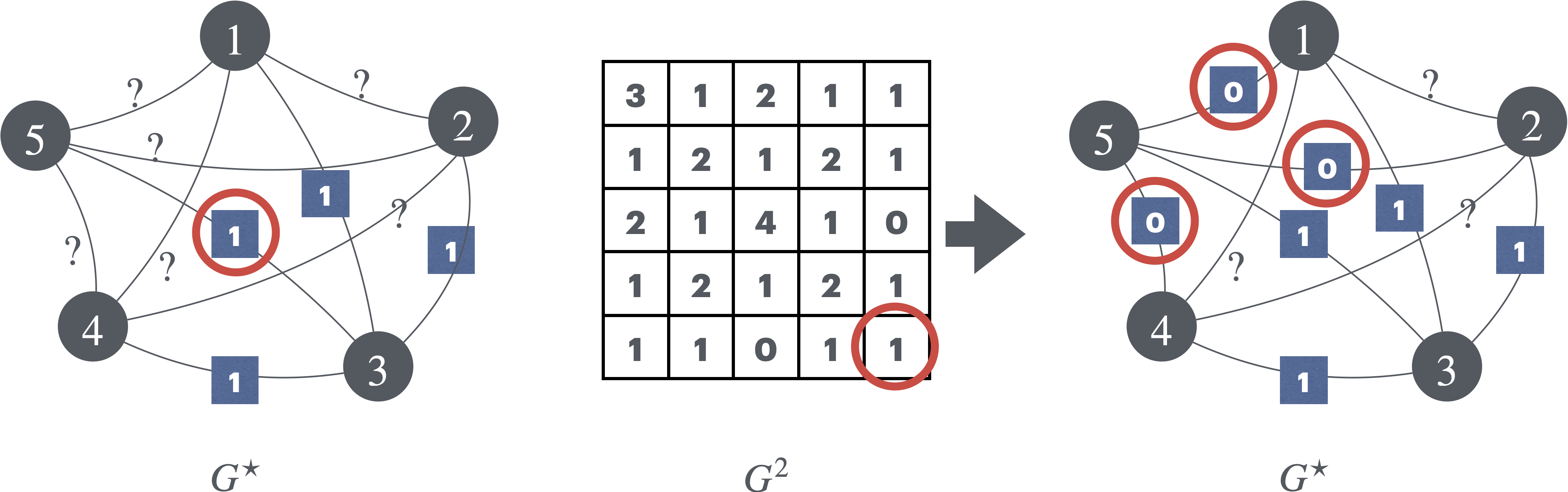}
    \caption{\DegreeMatchingAttack. The degree of vertex 5 (1) is equal to the number of its currently known neighbors in $\Gstar$. Thus all unknown edges in the neighborhood of 5 cannot exist.}
    \label{fig:degree-matching}
\end{figure}

\paragraph{\DegreeMatchingAttack}
From Proposition ~\ref{prop:2}, we derive Algorithm \ref{alg:degree-matching} which identifies the vertices whose neighborhoods have been completely recovered. This attack establishes the non-existence of edges in $G$ and puts 0's in $\Gstar$ by comparing the degrees of vertices on $\Gstar$ and $G$. An example of such inference is depicted in Figure ~\ref{fig:degree-matching}.

\begin{algorithm}
    \begin{algorithmic}[1]
    \caption{Identifies absent edges in $G$ based on the degrees of vertices in $G$ and $\Gstar$}
    \label{alg:degree-matching}
    \Require Partial reconstruction $G^{\star}$, common neighbors matrix $G^2$
    \Function{\DegreeMatchingAttack}{$\Gstar, G^2$}
    \For{$u \in \V$}
        \If{$G^2(u, u) = |\Gamma^{\star}(u)|$}

        \Comment{ \small The neighborhood of $u$ is complete}
            \For{$v \in \V \backslash \Gamma^{\star}(u)$} 
                \State $\Gstar(u, v) \leftarrow 0$ \Comment{ \small Update $\Gstar$}
            \EndFor
        \EndIf
    \EndFor
    \State \Return $\Gstar$
    \EndFunction
\end{algorithmic}
\end{algorithm}

\paragraph{\NeighborMatchingAttack}
The previous attack can also be applied to elements other than the diagonal of $G^2$, namely $G^2(u, v), u \neq v$. The intuition behind this reconstruction is the same: if the number of common neighbors between two vertices $u, v$ is the same on $\Gstar$ and $G$, it must be that there is no common neighbor on $G$ other than the ones known in $\Gstar$. This allows to insert 0s in $\Gstar$ by looking at the known neighbors. 

To see this, we focus on the special case of zero values in $G^2$. Such a value reflects the absence of common neighbors between the corresponding vertices: $\Gamma(u) \cap \Gamma(v) = \emptyset$. Of course, since the edges of $\Gstar$ are included in $G$, we also have $\Gamma^{\star}(u) \cap \Gamma^{\star}(v) = \emptyset$. More precisely : 
$$G^2(u,v) = 0 \implies [ \forall w \in \Gamma^{\star}(u), (w, v) \notin \E] $$
  
 (similarly with $u$ and $v$ interchanged).

 From the previous observation,  we devise the\\
 {\NeighborMatchingAttack}, depicted in Algorithm ~\ref{alg:neighbor-matching}.

\begin{algorithm}
\caption{ Identifies absent edges in $\Gstar$ based on the numbers of common neighbors in $G$.}
\label{alg:neighbor-matching}
\begin{algorithmic}[1]
\Require Reconstructed graph $\Gstar$, common neighbors matrix $G^2$

\Function{\NeighborMatchingAttack}{$\Gstar,  G^2$}
\For{$(u, v)  \in\V^2$ }%

    \If{$G^2(u, v) = |\Gamma^{\star}(u) \cap \Gamma^{\star}(v)|$}
      \\
      \Comment{The common neighborhood of $u$ and $v$ is complete}
      \For{$ w \in \Gamma^{\star}(u) \backslash (\Gamma^{\star}(u) \cap \Gamma^{\star}(v))$} 
      
          \State $G^*(w, v) \leftarrow 0$ \Comment{Update $\Gstar$}
      \EndFor
      \For{$w \in \Gamma^{\star}(v) \backslash (\Gamma^{\star}(u) \cap \Gamma^{\star}(v))$}
          \State $\Gstar(w, u) \leftarrow 0$ \Comment{ Update $\Gstar$}
      \EndFor
    \EndIf
\EndFor

\State \Return $\Gstar$
\EndFunction
\end{algorithmic}
\end{algorithm}

\paragraph{\DegreeCompletionAttack }

if $u$ is missing $k$ edges in $\Gstar$ to have a certain degree given by $G^2(u, u)$ and its neighborhood also has $k$  unknown edges, then all the unknown edges do exist.

\begin{figure}
    \centering
    \includegraphics[width=0.7\linewidth]{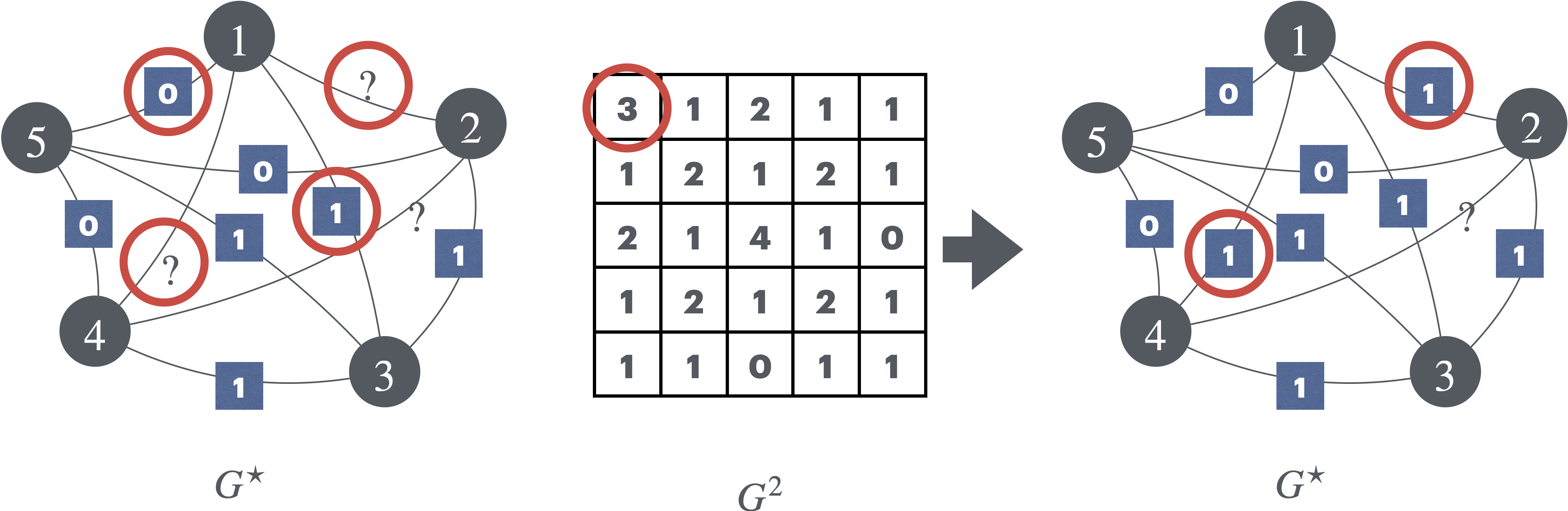}
    \caption{\DegreeCompletionAttack. Vertex 1 has one known neighbor which is 3, and two unknown edges. Also, it should have 3 neighbors in total. Therefore all its unknown edges do exist in $G$.}
    \label{fig:degree-completion}
\end{figure}

This inspired the {\DegreeCompletionAttack} presented in Algorithm \ref{alg:degree-completion} and depicted in Figure ~\ref{fig:degree-completion}
\begin{algorithm}
\caption{Identifies existing edges by comparing the neighborhoods of vertices on $G$ and $\Gstar$}
\label{alg:degree-completion}
    \begin{algorithmic}[1]
    \Require Reconstructed graph $\Gstar$, common neighbors matrix $G^2$
    \Function{\DegreeCompletionAttack}{$\Gstar, G^2$}
        \For{$u \in \V$}
            \State $ V \leftarrow \{v \in \V, \Gstar(u, v) = ? \}$
            \If{$G^2(u, u) = |\Gamma^{\star}(u)| + |V|$}
                \For{$v \in V$}
                    \State $\Gstar(u, v) \leftarrow 1$ \Comment{Update $\Gstar$}
                \EndFor
            \EndIf
         \EndFor  
     \State \Return $\Gstar$
     \EndFunction
\end{algorithmic}
\end{algorithm}

\paragraph{\NeighborCompletionAttack}
The {\DegreeCompletionAttack} can be extended to adapt it to the other information present in $G^2$, as presented in Algorithm \ref{alg:neighbor-completion}.
If $u, v$ have $m$ common neighbors on $\Gstar$ and $m+k$ on $G$, and $u$  has exactly $k$ unknown edges in its neighborhood, then it must be that all $k$ potential neighbors of $u$ are actually common neighbors of $u$ and $v$ in $G$.

\begin{algorithm}
\caption{ Identifies present edges by comparing the common neighbors of pairs of vertices on $G$ and $\Gstar$.}
\label{alg:neighbor-completion}
\begin{algorithmic}[1]
\Require Reconstructed graph $\Gstar$, common neighbors matrix $G^2$

\Function{\NeighborCompletionAttack}{$\Gstar, G^2$}
\For{$(u, v) \in\V^2 $ }%
        \State $ U \leftarrow \{w \in \V, \Gstar(u, w) = ? \}$
        \State $ V \leftarrow \{v \in \V, \Gstar(v, w) = ? \}$

        \If{$G^2(u, v) = |\Gamma^{\star}(u)| + |U|$}
            \For{$w \in U$}
                \State $\Gstar(u, w) \leftarrow 1$
                \State $\Gstar(v, w) \leftarrow 1$
            \EndFor
        \EndIf
        \If{$\Gsquare(u, v) = |\Gamma^{\star}(v)| + |V|$}
            \For{$w \in V$}
                \State $\Gstar(u, w) \leftarrow 1$
                \State $\Gstar(v, w) \leftarrow 1$
            \EndFor
        \EndIf
\EndFor

\State \Return $\Gstar$
\EndFunction
\end{algorithmic}
\end{algorithm}

\begin{figure}
    \centering
    \includegraphics[width=0.7\linewidth]{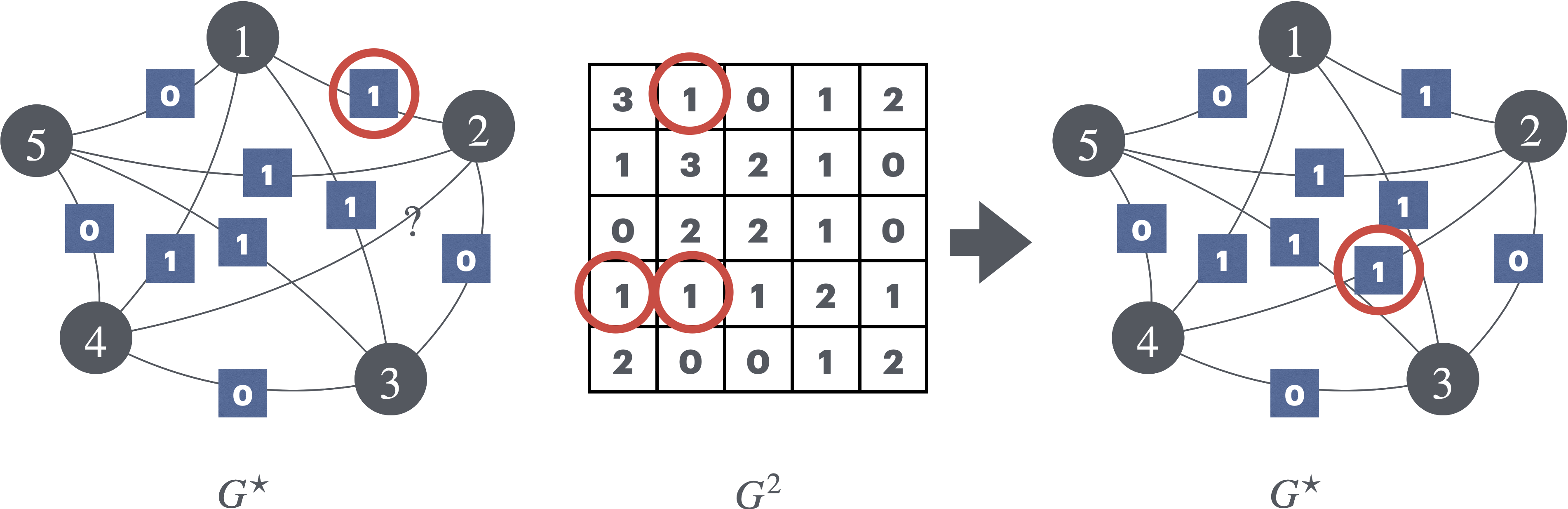}
    \caption{ \TriangleAttack.  Edge (1, 2) is known. Also, vertices 1 and 2 have one common neighbor. Then, there exists one triangle containing 1 and 2. Moreover, 4 is the only vertex that has common neighbors with both 1 and 2. Therefore 4 is the vertex forming a triangle with 1 and 2. }
    \label{fig:triangle}
\end{figure}

\paragraph{\TriangleAttack}
Proposition ~\ref{prop:triangle} allows the identification of triangles from the known edges in $G$, as well as $G^2$. Algorithm ~\ref{alg:triangle}, also illustrated in Figure ~\ref{fig:triangle}, identifies triangles by looking for vertices that share common neighbors with both vertices of an existing edge $(u, v)$.

\begin{algorithm}
    \caption{Identifies triangles in  $G$.}
    \label{alg:triangle}
    \begin{algorithmic}[1]
        \Require Reconstructed graph $\Gstar$, common neighbors matrix $G^2$

        \Function{\TriangleAttack}{$\Gstar, G^2$}
        \State $ \textit{edges} \gets \{(u, v), \Gstar(u, v) = 1 \}$ \Comment{Extract edges}
        \ForAll{$(u, v) \in \textit{edges}$}
    
            \State $ S_u = \{w: G^2(u, w) > 0 \text{ and } u \neq w \} $
              \State $ S_v = \{w: G^2(v, w) > 0 \text{ and } v \neq w \} $
              
              \Comment{Vertices that share common neighbors with $u$ and $v$}
            \State $S = S_u  \cap S_v$ 
            \If{$|S| = G^2(u, v)$ }
                \ForAll{$w \in S$}
                        \State $\Gstar(u, w) \gets 1$
                        \State $\Gstar(v, w) \gets 1$
                \EndFor
             \EndIf
         \EndFor
         \State \Return $\Gstar$
        \EndFunction
    \end{algorithmic}
\end{algorithm}

\begin{figure}
    \centering
    \includegraphics[width=0.7\linewidth]{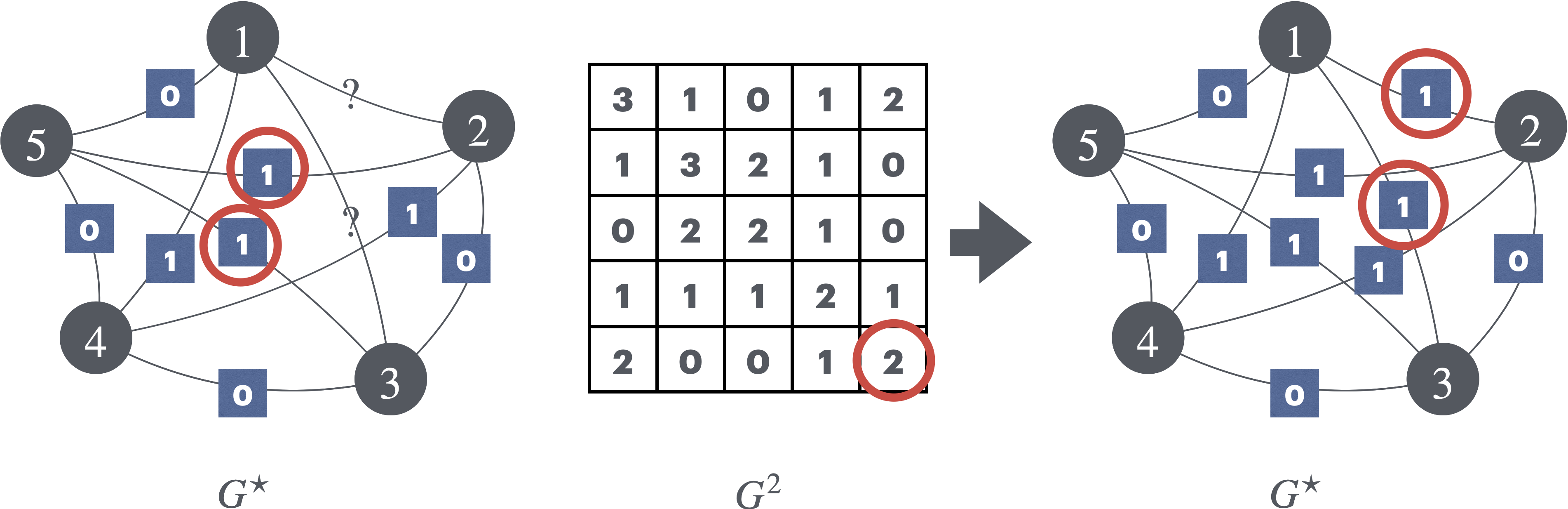}
    \caption{\BicliqueAttack. The neighbors of vertex 5 (which are 2 and 3) are all known, because its degree on $\Gstar$ is equal to $\Gsquare(5, 5)$. Therefore, vertex 1, which has two common neighbors with 5, is also connected to 2 and 3.}
    \label{fig:bi-clique}
\end{figure}

\paragraph{\BicliqueAttack}
When the neighborhood of a certain vertex $u$ is completely known, one can infer additional information about connections between other vertices and $u$. For instance, if all the $\Gsquare(u, u)$ neighbors of $u$ are known, and $v$ has $\Gsquare(u, u)$ common neighbors with $u$, then all the neighbors of $u$ are also neighbors of $v$. Inversely, if the number of missing neighbors for $v$ is equal to the number of missing common neighbors between $v$ and $u$, then all the missing neighbors of $v$ are among the neighbors of $u$. Therefore, $v$ does not have an edge with all the other vertices in the graph. Figure ~\ref{fig:bi-clique} gives an illustration of this inference, also detailed in Algorithm ~\ref{alg:bi-clique}.

\begin{algorithm}
\begin{algorithmic}[1]
        \caption{Identifies bi-cliques and their complements $G$.}
        \label{alg:bi-clique}
        \Require Reconstructed graph $\Gstar$, common neighbors matrix $\Gsquare$
        \Function{\BicliqueAttack}{$\Gstar, \Gsquare$ }
            \State $U \gets \{u, G^2(u, u) = |\Gamma^{\star}(u)|\}$ \Comment{Nodes with satisfied degree}
            \ForAll{$u \in U$}
                \ForAll{$v \in \V$}
                    \If{$\Gsquare(u, v) = \Gsquare(u, u)$} \Comment{Bi-clique found}
                         \ForAll{$w \in \Gamma^{\star}(u)$ }
                            \State $\Gstar(v, w) \gets 1$
                        \EndFor
               
                    \Else
                        \State $m_G = \Gsquare(v, v) - |\Gamma^{\star}(v)|$
                        \State $m_{\Gsquare} = \Gsquare(u, v) - |\Gamma^{\star}(u) \cap \Gamma^*(v)|$

                        \If{$m_G = m_{G^2}$}
                            \ForAll{$w \in \V \backslash \Gamma^{\star}(u)$}
                                \State $\Gstar(v, w) \gets 0$
                            \EndFor
                        \EndIf
                  \EndIf
                \EndFor
            \EndFor
            \State \Return $\Gstar$
        \EndFunction
\end{algorithmic}
\end{algorithm}

\paragraph{Composing the attacks}
Our topological attacks come in two flavors. Some of them, like the {\DegreeMatchingAttack} and {\NeighborMatchingAttack} identify non-existing edges in the graph, changing $?$s into 0s in $\Gstar$. Inversely, attacks such as the {\DegreeCompletionAttack} and {\NeighborCompletionAttack} identify existing edges, changing $?$s into 1s states.

Since the changes made by one type of attack can influence the output of the other, the process needs to be repeated until neither of the algorithms updates $\Gstar$. Thus, we derive the iterative procedure depicted in Algorithm \ref{alg:deterministic-attacks} for the execution topological attacks.

\begin{algorithm}
\begin{algorithmic}[1]
\caption{ToplogicalAttack}
\label{alg:deterministic-attacks}
\Require Lists of edges and non-edges $\E_1, \E_0$, common neighbors matrix $G^2$

\Function{TopologicalAttack}{$\E_1, \E_0, G^2$}
\State $\Gstar \leftarrow$ \Call{Init}{$\E_1, \E_0$}  \Comment{Copy 1s and 0s into $\Gstar$}
    \While{$\Gstar$ has been updated} 
        \State $\Gstar \leftarrow $  \Call{\DegreeCombinationAttack}{$\Gstar, \Gsquare $}
        \State $\Gstar \leftarrow $  \Call{\DegreeMatchingAttack}{$\Gstar, \Gsquare $}
        \State $\Gstar \leftarrow $  \Call{\NeighborMatchingAttack}{$\Gstar, \Gsquare $}
        \State $\Gstar \leftarrow $  \Call{\DegreeCompletionAttack}{$\Gstar, \Gsquare $}
        \State $\Gstar \leftarrow $  \Call{\NeighborCompletionAttack}{$\Gstar, \Gsquare $}
        \State $\Gstar \leftarrow $  \Call{\TriangleAttack}{$\Gstar, \Gsquare $}
        \State $\Gstar \leftarrow $  \Call{\BicliqueAttack}{$\Gstar, \Gsquare$}
    \EndWhile
    \State \Return  $\Gstar$
\EndFunction
\end{algorithmic}    
\end{algorithm}

\subsection{Spectral attack}
Hereafter, we introduce algorithms for the reconstruction of $G$ based on its spectrum. 

 Our solution revisits the greedy sign assignment algorithm of \cite{Erdös12}, and improves it by taking into account an additional constraint related to the partial knowledge of the graph $G$ we want to reconstruct. The idea is to iteratively find the best sign assignment for each singular value, starting from the first. 
 Let us first denote by  $\E^{\star}_0$ the 0s in $\Gstar$ so far, $\E^{\star}_1$ the 1s, and $\E^{\star} = \E^{\star}_1 \cup \E^{\star}_0$. For each singular value, the best sign is the one which satisfies the following two constraints :

\begin{itemize} 
    \item \textbf{Closeness to a binary matrix : } similar to \cite{Erdös12}, we impose that the reconstructed matrix be a binary one.
    \item  \textbf{Closeness to the known adjacency information : } to account for the already known information on the graph, we impose that our reconstruction be close to it. Namely, for a given potential edge $(u, v)$, our reconstructed adjacency matrix should contain a value of 1 ( respectively 0 ) at index (u, v) if $(u, v) \in \E^{\star}_1 \text{ (respectively } \E^{\star}_0 \text{)}$.
\end{itemize}

\begin{algorithm}
\caption{SpectralAttack}
\label{alg:SpectralAttack}
\begin{algorithmic}[1]
\Require Partial reconstructed graph $\Gstar$, SVD of $\Gsquare$ ($U, \Lambda, V$  such that $G^2 = U \Lambda V$), $\alpha, \beta$ weighting factors.
\Function{SpectralAttack}{$\Gstar, U, \Lambda, V, \alpha, \beta$ }
\State $M_0 \gets \mathbf{0}^{n\times n}$ 
\State $\E^{\star} \gets \{(i,j) \mid \Gstar(i,j) \in \{0, 1\}\}$ \Comment{Extract edges and non-edges from $\Gstar$}
\For{$i = 1$ to $|\V|$}

    \State $M^+_i \gets M_{i-1} + U(:,i) \cdot \sqrt{\Lambda(i, i)} \cdot V(:,i)^T$
    \State $M^-_i \gets M_{i-1} - U(:,i) \cdot \sqrt{\Lambda(i, i)} \cdot V(:,i)^T$
    \State $d^+ \gets \alpha \norm{M^+_i - BM^+_i}_F + \beta \norm{M^+_i\vert_{\E^{\star}} - \Gstar\vert_{\E^{\star}}}_F$
    \State $d^- \gets \alpha \norm{M^-_i - BM^-_i}_F + \beta \norm{M^-_i\vert_{\E^{\star}} - \Gstar\vert_{\E^{\star}}}_F$
    \If{$d^+<d^-$}
        \State $M_i \gets M^+_i$
    \Else
        \State $M_i \gets M^-_i$
    \EndIf
\EndFor
\State \Return $BM$
\EndFunction
\end{algorithmic}
\end{algorithm}

\begin{algorithm}
    \caption{TargetedErrorForgetting}
    \label{alg:SanityCheck}
    \begin{algorithmic}[1]
        \Require Graph reconstructed by spectral attack $\Gstar_s$, Graph reconstructed by topological attacks $\Gstar_t$,  common neighbors matrix $\Gsquare$
        \Function{TargetedErrorForgetting}{$\Gstar_s, \Gstar_t, \Gsquare$}
        
            \Comment{Common neighbors matrix on the reconstructed graph}
            \State $G^{\star2}_s \gets \Gstar_s \cdot \Gstar_s$ 
            \For{$(u, v) \in \V^2$}
                
                \Comment{ Incorrect common neighbors}
                    \If{$\Gsquare(u, v) \neq G^{\star2}_s(u, v)$} 
                        \For{$w \in \V$}
                                \State $\Gstar_s(u, w) \gets \Gstar_t(u, w)$
                                \State $\Gstar_s(v, w) \gets \Gstar_t(v, w)$
                        \EndFor
                    \EndIf
                \EndFor
            \State \Return $\Gstar_s$
        \EndFunction
    \end{algorithmic}
\end{algorithm}

We formalize the above constraints as a minimization problem, where for the \textit{i-th} singular value, we choose the sign $s \in \{+, -\}$ that minimizes the joint distance with our objectives :

$$
 min_{s \in \{+,- \}} \alpha \norm{M^s - BM^s}_F + \beta \norm{M^s\vert_{\E^*} - \Gstar \vert_{\E^*}}_F
$$

where $M^s$ denotes the reconstructed matrix computed with singular value $s\sqrt{\Lambda(i, i)}$.  $BM^s$ is the binary version of the reconstructed matrix $M^s$. Precisely, for each row $i$ and column $j$ of $M^s$,
\[
BM^s(i, j) =
\begin{cases} 
1 & \text{if } M^s(i, j) > t, \\
0 & \text{otherwise,}
\end{cases}
\]
where $t$ is a chosen binarization threshold.
The $\alpha$ term weights in for the reconstruction of a binary matrix, and the $\beta$ term allows for the reconstruction of a matrix close to the known information in $\Gstar$.
The $\alpha$ and $\beta$ weighting factors allow a trade-off between the two constraints. This is especially suitable for the case when we already know a big part of the matrix, as we can boost the second term to benefit from that partial knowledge.
$M^s{\vert{_{\E^{\star}}}}$ denotes $M^s$ restricted to the indices (rows and columns) contained in $\E^{\star}$.

  Our algorithm for the spectral reconstruction of $G$, denoted \textsc{SpectralAttack} is described in Algorithm \ref{alg:SpectralAttack}.

\subsection{Targeted error forgetting}
Very often, the spectral attack reconstructs $G$ with error. That is, some 1s are mistaken for 0s, and inversely some 0s for 1s. 
While correcting errors in the reconstruction is not straightforward, we can leverage our knowledge of $\Gsquare$ to  detect some of them, as depicted in Algorithm \ref{alg:SanityCheck}. Indeed, consider two vertices $u, v$ that do not have the right number of common neighbors in the reconstruction. Then, it must be that there is an error either in the neighborhood of $u$, in the neighborhood of $v$ or both.
Therefore, in such cases we can rollback the updates of the spectral attack, and put ?s in the neighborhood of $u$ and $v$.
This heuristic cannot delete all the errors in the reconstruction (since $u$ and $v$ could have the right number of common neighbors, but the wrong common neighbors), but we experimentally find that is allows to curate most of them, at the expense of some false positives.

\subsection{Co-square graph instantiation}
\label{sec:cosquare-instanciation}
\begin{figure}
    \centering
    \includegraphics[width=0.7\linewidth]{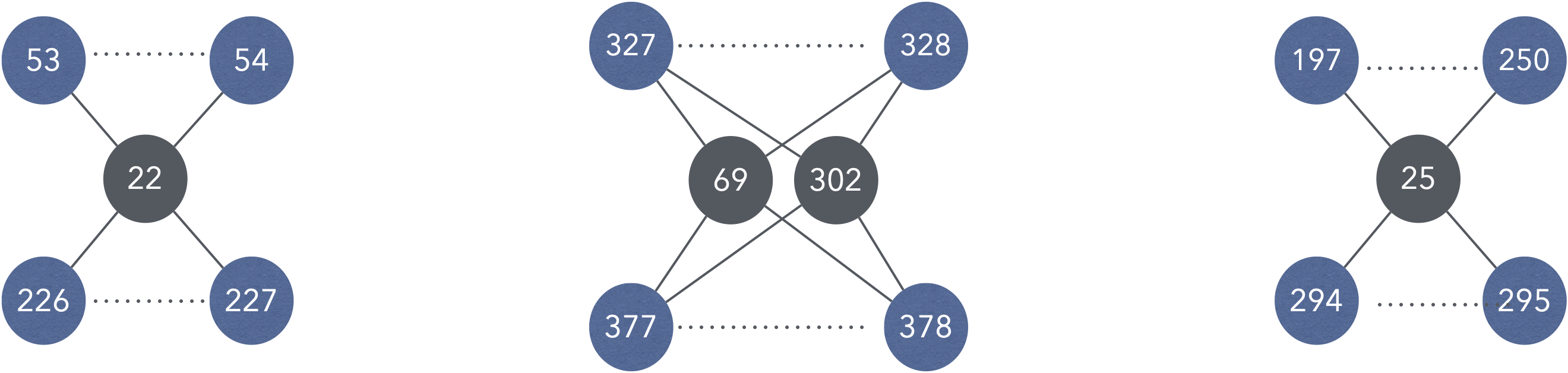}
    \caption{Co-square sub-graphs in Netscience. The plain lines are reconstructed edges, the dotted lines are the missing edges in the reconstruction. For each component, adding the two missing edges between any combination of two of the four blue vertices yields a graph with equal matrix of common neighbors. }
    \label{fig:strucural-equivalence}
\end{figure}

Because of the possible existence of a co-square graph, some entries in $G^2$ can be achieved with different attribution of edges in $\Gstar$.  
An example of such a scenario, extracted from Netscience, is depicted in Figure ~\ref{fig:strucural-equivalence}--the number labels are the ones of Netscience. 

 \begin{figure*}
    \centering
    \includegraphics[width=\linewidth]{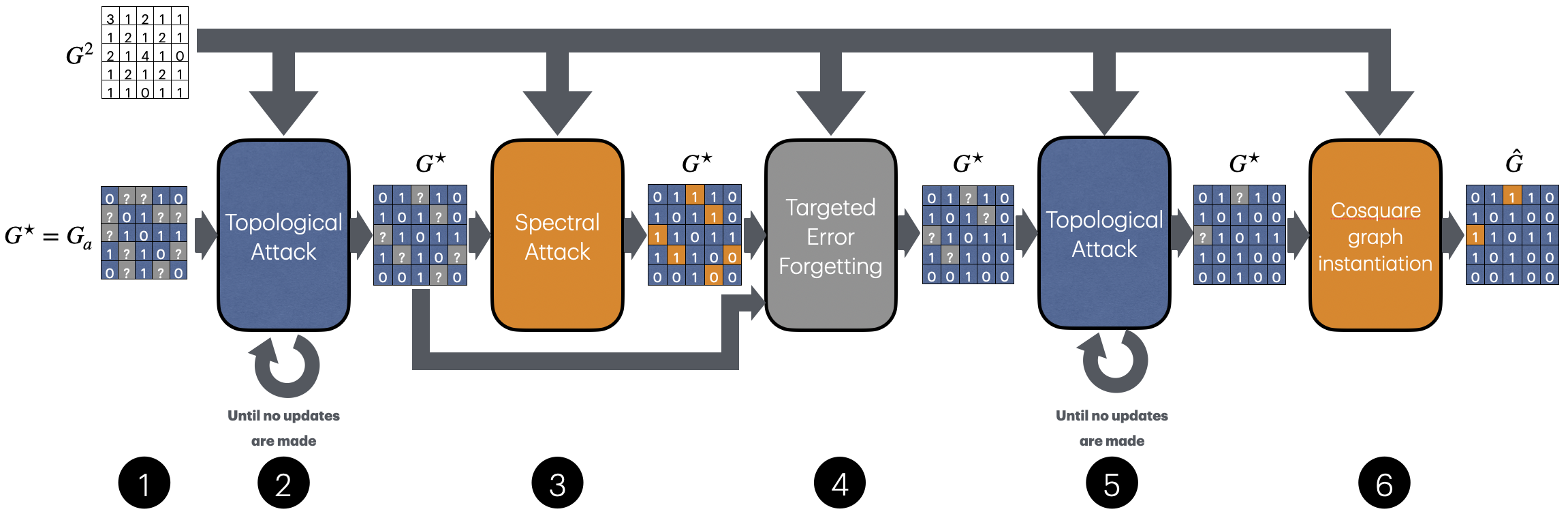}
    \caption{GRAND}
    \label{fig:pipeline}
\end{figure*}

For each separate component, the position of dotted edges w.r.t. the blue vertices can be in three possible ways to yield the values in $G^2$:  as they are pictured, or they could be set vertically or in diagonal. For example for the left example, the edges can be replaced by edges $(53,226)$ and $(54,227)$ or by  $(53,227)$ and $(54,226)$.
Interestingly, the first graph of 5 vertices (hence detached from Netscience) and its two co-square graphs (which we just described) are the smallest co-square graphs. They are isomorphic, contrarily to the first co-square graphs that we presented in Section~\ref{sec:non-unicity}, which are the smallest non isomorphic co-square graphs.

In such cases, a prior knowledge on these difficult components of $G$ is essential to reconstruct it, as there are a certain number of graphs that have the same $\Gsquare$. 

In our method, when no prior knowledge is given, or when it does not contain any information about the co-square-inducing vertices, we instantiate one co-square of the target graph after identifying the co-square-inducing vertices based on $\Gsquare$.

 \subsection{Pipeline}
 Our complete attack pipeline, depicted in Figure \ref{fig:pipeline} unfolds as follows:
 \begin{enumerate}
     \item We initialize $\Gstar$ with the known information $\E_0, \E_1$. When $\E_0 = \E_1 = \emptyset$, the adjacency matrix of this graph contains 0s on the main diagonal (because we know beforehand that there is no loop in the graph), and $?$s everywhere else;
     \item We run the topological attacks until no new information is discovered on $G$.
     \item We then pass the partial information gathered so far $\Gstar$ to the spectral attack, which optimizes the reconstruction to be close to the known edges and non-edges we learned.
     \item Since the spectral method might introduce errors, we perform a targeted error forgetting that removes the adjacency information of vertices in $\Gstar$ based on $G^{\star2}$. 
     \item The error forgetting introduces $?$s in $\Gstar$. To recover values, we run another round of topological attacks.
     \item To identify edges that were not recovered so far because of the co-squareness, a co-square subgraph instantiation is performed.
 \end{enumerate}

\section{Experiments and analysis}
\label{sec:experiments}
In the following, we present the performance of our algorithms evaluated on various real-world datasets. We compare our reconstruction capability with the one reached by \cite{Erdös12}, while taking into account the prior knowledge $\E_0, \E_1$. We simulate this prior knowledge by a uniform sampling of the matrix $G$, with a proportion of $\rho$. In other words, the attacker is allowed to know a proportion $\rho$ of 1s (edges) and 0s (non edges) from $G$.

The datasets used in our experiments are presented in Table ~\ref{tab:datasets}. 

\begin{table}
    \centering
    \begin{tabular}{|c|c|c|c|c|c|}
        \hline
        Dataset & $|\V|$ & $|\E|$ & Density & Domain \\  
        \hline
        Netscience \cite{networkrepo} & 379 & 914 & 0.0127 & Co-authorship \\
        Bio-diseasome \cite{networkrepo} & 516 & 1188 & 0.0089 & Human diseases  \\
        Polblogs \cite{li2020deeprobust} & 1490 & 16715 & 0.0151 & USA Political blogs  \\
        Cora \cite{li2020deeprobust} & 2485 & 5069 & 0.0016 & Citations\\
        \hline
    \end{tabular}
    \caption{Datasets}
    \label{tab:datasets}
\end{table}
We evaluate the reconstruction performance for both of the adversary models detailed in Section \ref{sec:problem-statement}. Section \ref{sec:ga-empty} presents the results obtained when no prior knowledge is given to the attacker that is, in the presence of an unknowledgeable attacker. Then we explore in Section \ref{sec:ga-non-empty} the reconstruction performance with a varying amount of prior knowledge on $G$ given to the knowledgeable attacker. 

\paragraph{Evaluation metrics}
We consider four metrics to measure the reconstruction performance:
\begin{itemize}
    \item The False Positives Rate (FPR), which measures the amount of edges in the reconstructed graph that do not belong to the original;
    \item The False Negatives Rate (FNR), which is the amount of edges which belong to the original graph but were not recovered by the attack;
    \item The Relative Absolute Error (RAE). This measures the amount of wrong predictions (positives and negatives) in proportion of the number of edges in $G$ (also used in~\cite{Erdös12}).
    $$RAE = \frac{\norm{G-\hat{G}}_F}{\norm{G}_F}$$
    \item We introduce the \emph{Common Neighbors Error}, as a measure of error with regard to the common neighbors matrix. This measure serves as an evaluation metric that takes into account the co-square equivalence explained in Section~\ref{sec:non-unicity}. To a real attacker, it can also serve as an indicator of how close they are to the real graph without knowing it beforehand.
    $$CNE = \frac{\norm{\Gsquare-\hat{G}^2}_F}{\norm{\Gsquare}_F}$$
\end{itemize}

The parameters of the spectral attack are chosen as follows :
\begin{itemize}
    \item $\beta = \frac{2 \cdot |\E^*|}{|\V|^2}$. Since $\E^*$ contains the indices of the known information in the reconstructed matrix, this choice of $\beta$ ensures that the more we know about $G$, the more we use that information in our reconstruction. When $G$ is completely known, $\beta = 1$.
    \item $\alpha = 1-\beta$, to complement for the unknown information.
    \item $t = 0.5$. This choice acts as a middle ground for binarizing values between 0 and 1. In \cite{Erdös12}, Erd\H{o}s et al. also report this value as the best for their approach. Therefore, we use the same as theirs.
\end{itemize}

We present average values taken over 10 runs.

\subsection{Reconstruction by an unknowledgeable attacker}
\label{sec:ga-empty}
\begin{figure}[H]
    \centering
    \includegraphics[width=0.7\linewidth]{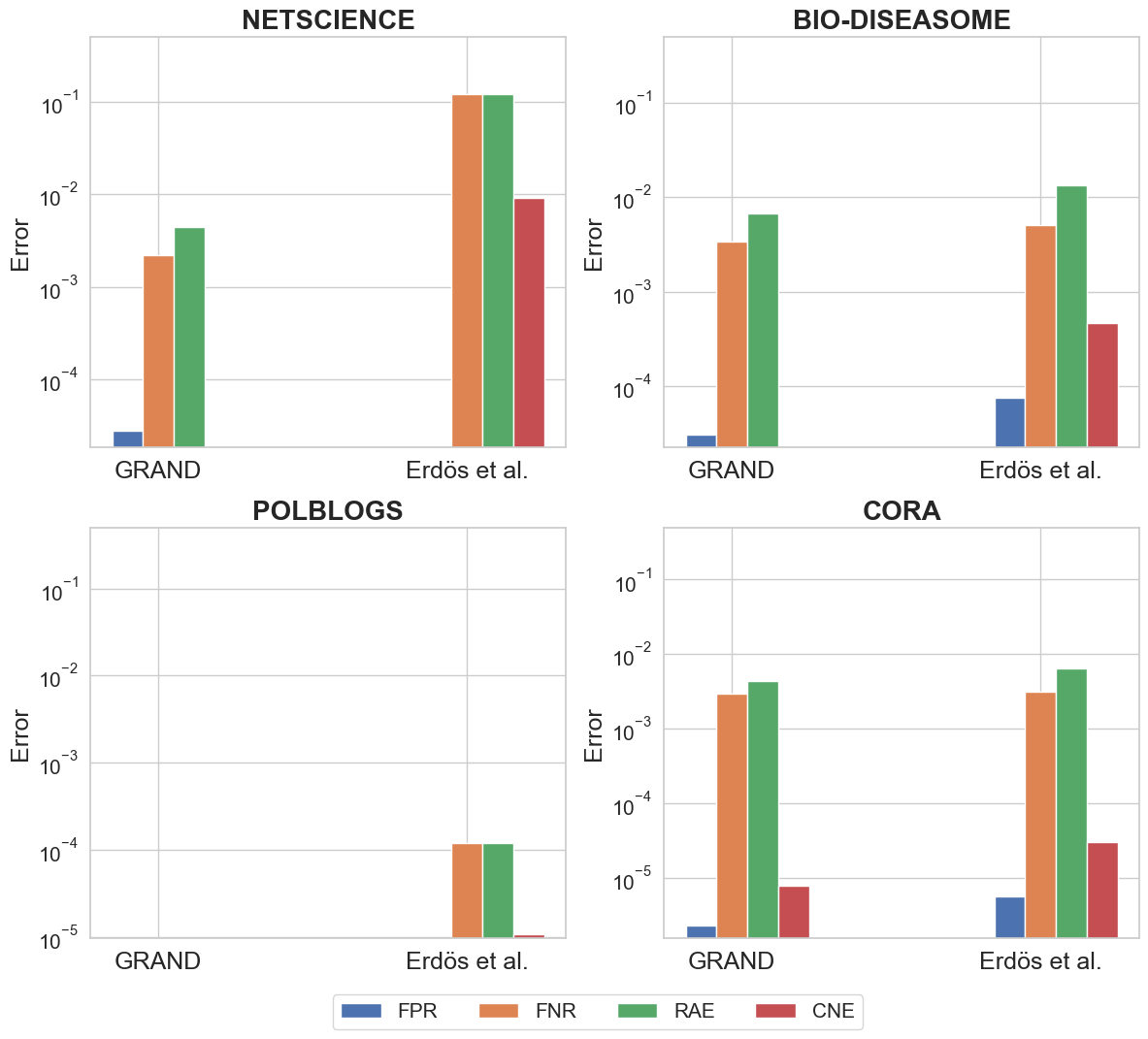}
    \caption{False Positives Rate, False Negatives Rate, Relative Absolute Error and Common Neighbors Error  with $\rho$=0.}
    \label{fig:errors-ga-empty}
\end{figure}

In this scenario, the attacker gets no prior knowledge. That is, all the edges in $\Gstar$ are labeled as ?s.
Figure \ref{fig:errors-ga-empty} presents the FPR, FNR, RAE and CNE obtained using GRAND in comparison to the method from \cite{Erdös12}. 
GRAND minimizes the false positives and and false negatives, leading to a reduced RAE in comparison to \cite{Erdös12}.
GRAND even achieves a perfect reconstruction on the Polblogs dataset.

\subsection{Reconstruction by a knowledgeable attacker}
\label{sec:ga-non-empty}
For this scenario, the common neighbors matrix $\Gsquare$ is passed to our attack pipeline, as well as $\Gstar$, initialized as described previously. However, it should be mentioned again that the approach of \cite{Erdös12} does not take into account a prior knowledge of the graph, and there is no straightforward way to include it either. In order to have a fair comparison with that approach, we consider in this scenario that when using the approach of \cite{Erdös12}, the attacker reconstructs a graph $\hat{G}$ that they overwrite using the prior knowledge. Namely, edges in $\E_0$ are removed from $\hat{G}$ and edges in $\E_1$ are added to $\hat{G}$ regardless of the initial value in $\hat{G}$. All other possible edges stay as they were in $\hat{G}$. We call this modified approach of \cite{Erdös12} the \emph{knowledgeable Erdös et al.}

\begin{figure}
    \centering
    \includegraphics[width=0.7\linewidth]{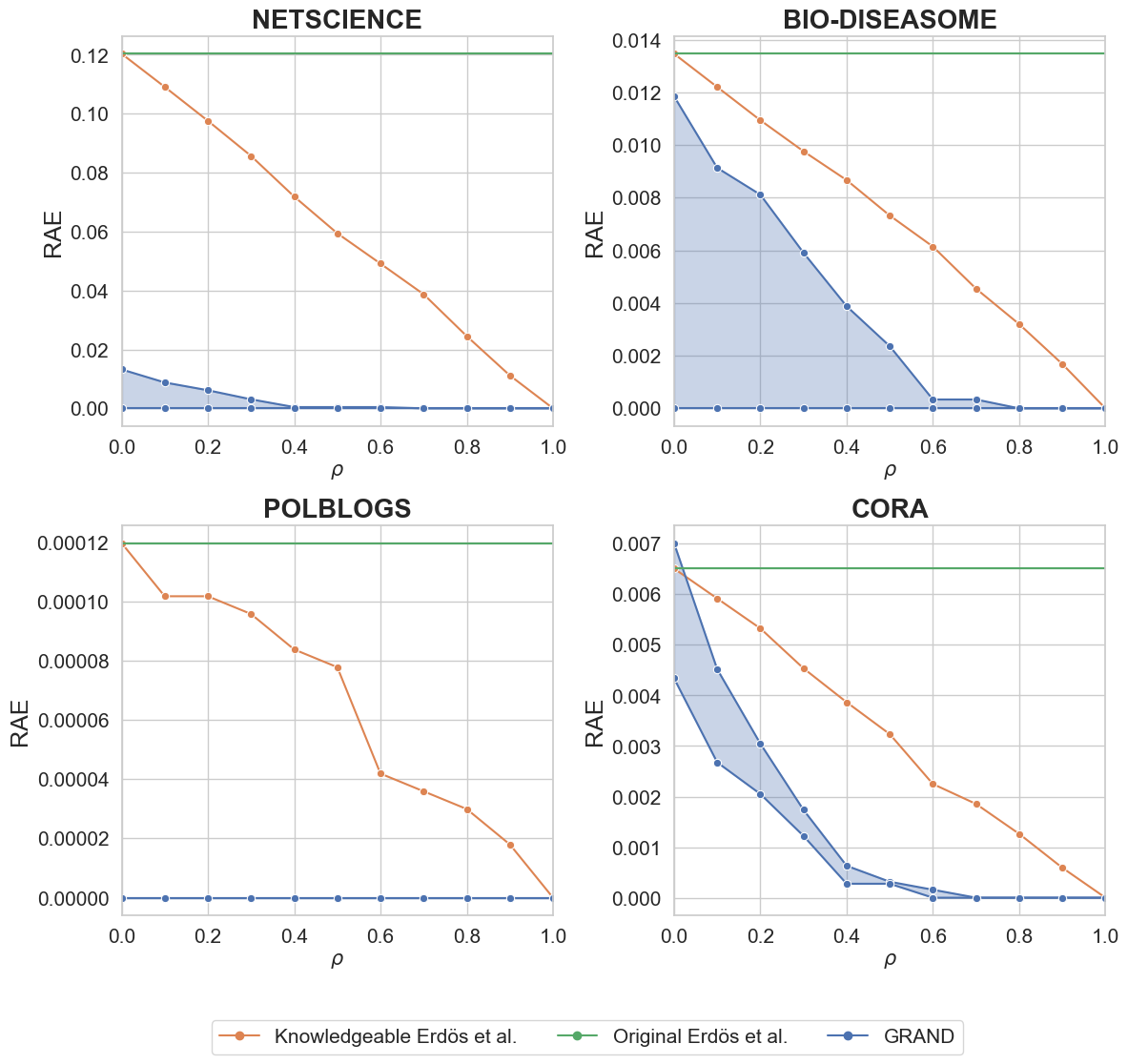}
    \caption{Relative Absolute Error (RAE) with respect to the amount of prior knowledge $\rho$}
    \label{fig:rae-ga-non-empty}
\end{figure}

\paragraph{From the perspective of $G$}
Figure ~\ref{fig:rae-ga-non-empty} presents the RAE of both approaches with respect to the amount of prior knowledge. The shade for GRAND denotes the maximum and minimum RAE obtained based on the instantiated co-square subgraph.
Both approaches benefit from the additional prior knowledge, and GRAND consistently reaches a lower RAE on all datasets, for all values of $\rho$.

\begin{figure}
     \centering
     \includegraphics[width=0.7\linewidth]{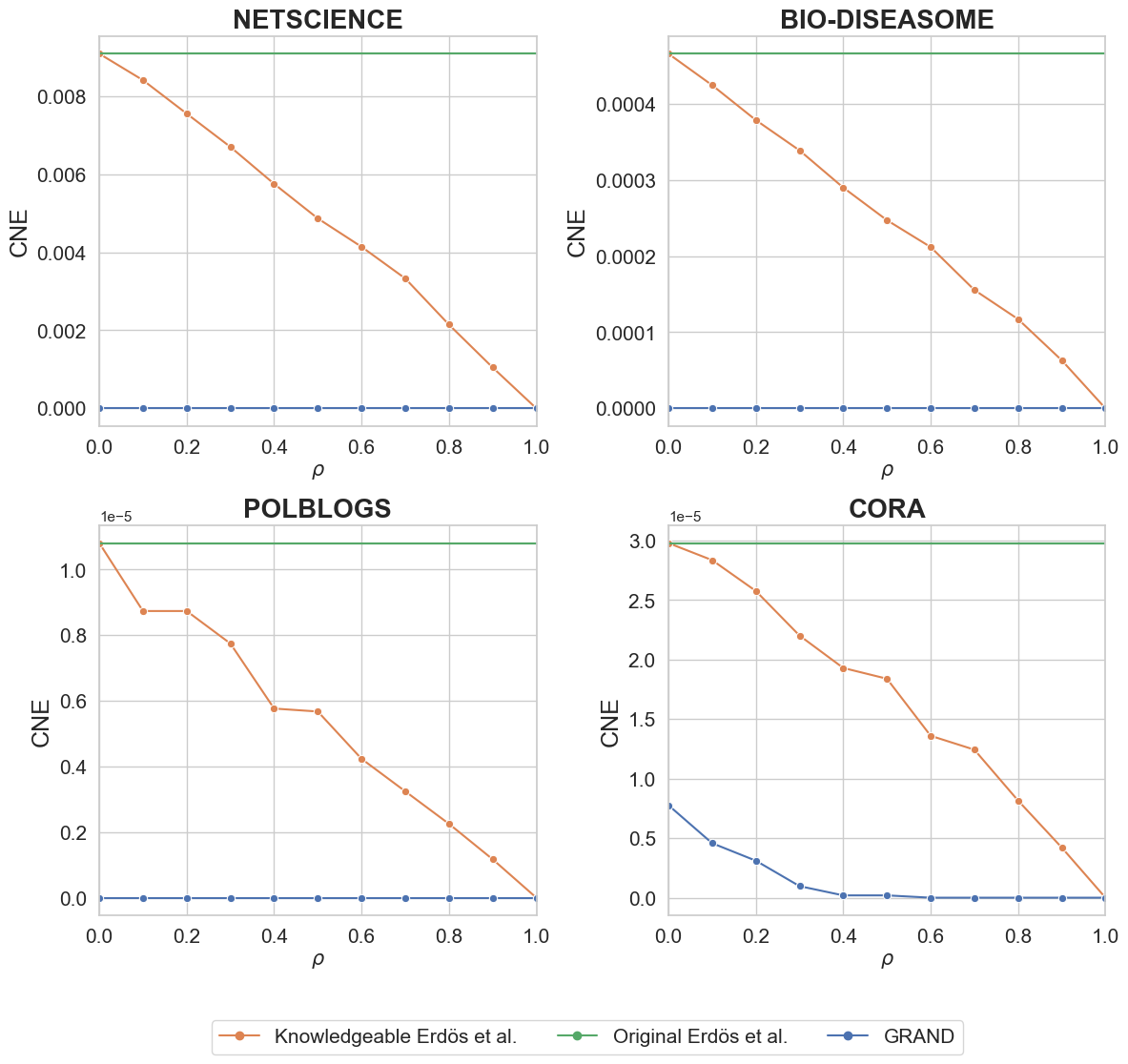}
     \caption{Common Neighbors Error with respect to the amount of prior knowledge $\rho$}
     \label{fig:cne}
 \end{figure}

\paragraph{From the perspective of $\Gsquare$}
The results in terms of CNE, presented in Figure ~\ref{fig:cne}, complement previous ones by alleviating the consideration for co-square graphs. Without any prior knowledge, GRAND reconstructs perfect co-square graphs of the target graph for Netscience, Polblogs and Bio-diseasome, bringing the CNE to 0. This shows that GRAND actually reaches the theoretical maximum reconstruction performance, since without prior knowledge the attacker cannot guess which co-square graph corresponds to the target graph.

\begin{figure}[H]
    \centering
    \includegraphics[width=0.7\linewidth]{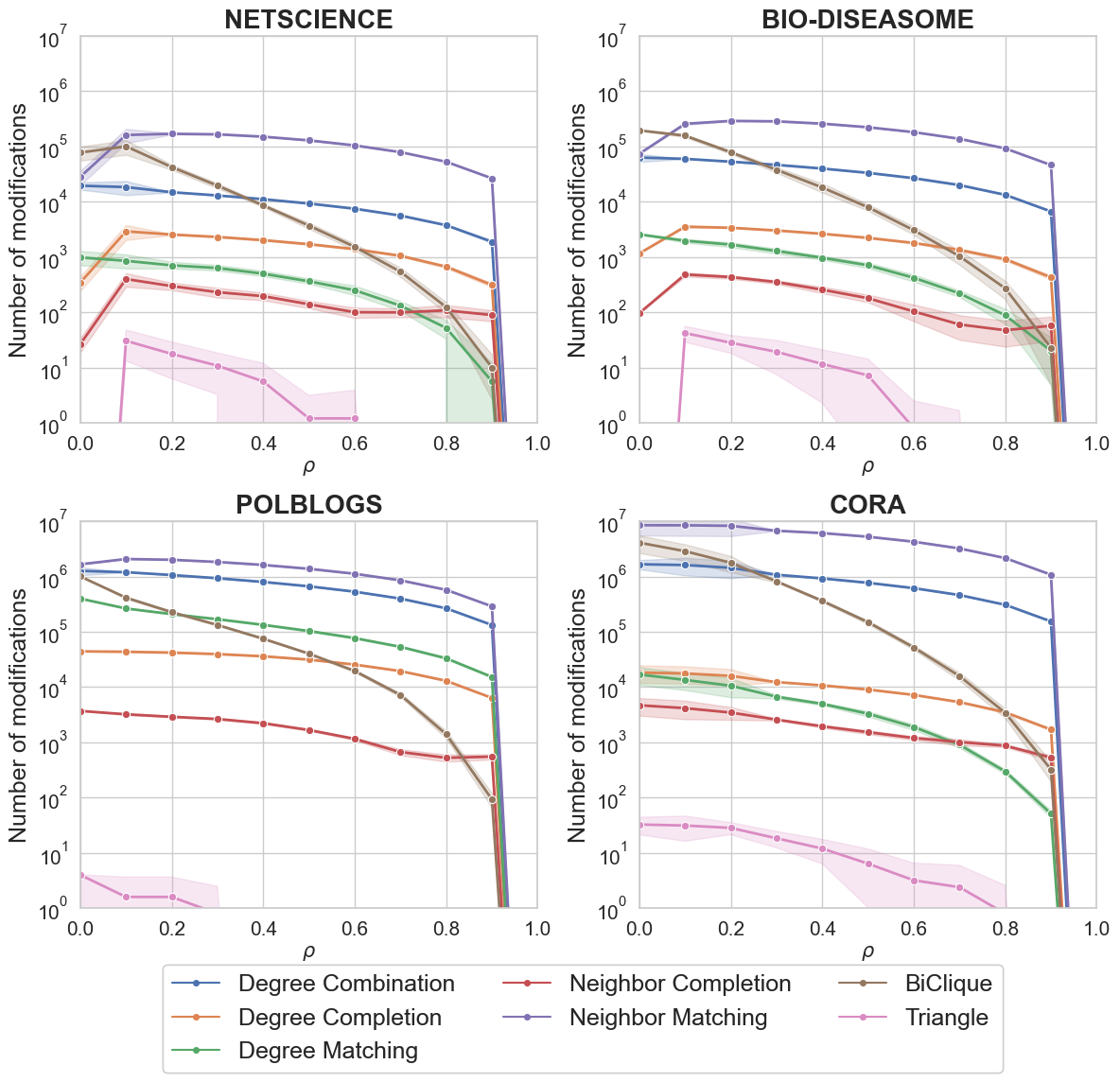}
    \caption{Number of modifications made by the topological attacks.}
    \label{fig:modifs-topological}
\end{figure}

\paragraph{More about topological attacks}
Figure ~\ref{fig:modifs-topological} showcases the importance of each topological attack in terms of numbers of modifications (changes form ? to 0 or 1) in the adjacency matrices of the reconstructed graphs. Since these matrices are very sparse, attacks that only find 1s (e.g. \TriangleAttack,  \NeighborCompletionAttack) tend to make less modifications in the graph. However, as pointed out before, the various algorithms complement each other by finding existing and non existing edges that allow subsequent inferences for the next iterations.

\section{Conclusion}
\label{sec:conclusion}

In this paper, we have introduced GRAND, a novel approach to 
reconstructing graphs from their neighborhood data given by the 
common neighbors matrix. This can have significant implications in real-world scenarios, such as revealing hidden connections in social networks or communication systems. For that, we took two angles of attack: the topological one and the spectral one. The topological angle allows us to infer certain information about the existence or non-existence of edges in the target graph based on the number of common neighbors of these vertices and on their degrees. We enumerate a list (which we believe is non-exhaustive) of properties that allow us to infer this information about the graph's structure. The spectral angle uses SVD decomposition by incorporating information from the topological attack, reconstructing iteratively the graph through its eigenvalues and minimizing the distance to the partially recovered graph.
We have shown, through multiple experiments on various datasets, that our approach is, for most of the cases, able to reconstruct graphs up to co-squarity, even when we don't have any prior knowledge about the target graph.

Along the way, we have defined co-squareness, a new notion of equivalence between graphs that is specific to the content of their common neighbor matrices, a property that is not shared by most isomorphic graphs. Moreover, to compare fairly a reconstructed graph to the one we are looking for, we have introduced an appropriate metric, which focuses on the values of the square matrix, since, in the absence of prior knowledge, graphs that are co-square  to the original one are as good as one can expect.

\paragraph{Future work} Given the vast diversity of graphs, it is very challenging to generalize any attack to all types of graphs. Therefore, one of the future directions of this work will be to extend the attack to directed graphs, bipartite graphs, and other types of graphs. Another direction will be to study the NP-hardness of the problem by reducing it to a known NP-hard problem.

\bibliographystyle{plainurl}
\bibliography{main.bib}

\begin{thebibliography}{10}

\bibitem{CryptoGraph}
Sofiane Azogagh, Zelma~Aubin Birba, S\'{e}bastien Gambs, and Marc-Olivier Killijian.
\newblock Crypto'graph: Leveraging privacy-preserving distributed link prediction for robust graph learning.
\newblock 2024.

\bibitem{bai2024characterizing}
Yandong Bai, Pedro~P Cort{\'e}s, Reza Naserasr, and Daniel~A Quiroz.
\newblock Characterizing and recognizing exact-distance squares of graphs.
\newblock {\em Discrete Mathematics}, 347(8):113493, 2024.

\bibitem{CHVETAL1980249}
V.~Chvétal.
\newblock Recognizing intersection patterns.
\newblock In M.~Deza and I.G. Rosenberg, editors, {\em Combinatorics 79 Part I}, volume~8 of {\em Annals of Discrete Mathematics}, pages 249--251. Elsevier, 1980.
\newblock URL: \url{https://www.sciencedirect.com/science/article/pii/S0167506008708835}, \href {https://doi.org/10.1016/S0167-5060(08)70883-5} {\path{doi:10.1016/S0167-5060(08)70883-5}}.

\bibitem{cloteaux16}
Brian Cloteaux.
\newblock Fast sequential creation of random realizations of degree sequences.
\newblock {\em Internet Mathematics}, 12(3):205--219, 2016.

\bibitem{daud2020applications}
Nur~Nasuha Daud, Siti~Hafizah Ab~Hamid, Muntadher Saadoon, Firdaus Sahran, and Nor~Badrul Anuar.
\newblock Applications of link prediction in social networks: A review.
\newblock {\em Journal of Network and Computer Applications}, 166:102716, 2020.

\bibitem{Demirag23}
Didem Demirag, Mina Namazi, Erman Ayday, and Jeremy Clark.
\newblock Privacy-preserving link prediction.
\newblock In Joaquin Garcia-Alfaro, Guillermo Navarro-Arribas, and Nicola Dragoni, editors, {\em Data Privacy Management, Cryptocurrencies and Blockchain Technology}, pages 35--50, Cham, 2023. Springer International Publishing.

\bibitem{eckart1936approximation}
Carl Eckart and Gale Young.
\newblock The approximation of one matrix by another of lower rank.
\newblock {\em Psychometrika}, 1(3):211--218, 1936.

\bibitem{Erdös12}
Dóra Erdös, Rainer Gemulla, and Evimaria Terzi.
\newblock Reconstructing graphs from neighborhood data.
\newblock In {\em 2012 IEEE 12th International Conference on Data Mining}, pages 231--240, 2012.
\newblock \href {https://doi.org/10.1109/ICDM.2012.154} {\path{doi:10.1109/ICDM.2012.154}}.

\bibitem{hasan2011survey}
Mohammad~Al Hasan and Mohammed~J Zaki.
\newblock A survey of link prediction in social networks.
\newblock {\em Social network data analytics}, pages 243--275, 2011.

\bibitem{horn2012matrix}
Roger~A Horn and Charles~R Johnson.
\newblock {\em Matrix analysis}.
\newblock Cambridge university press, 2012.

\bibitem{zoltan12}
Zolt{\'a}n Kir{\'a}ly.
\newblock Recognizing graphic degree sequences and generating all realizations.
\newblock {\em E{\"o}tv{\"o}s Lor{\'a}nd University, Tech. Rep. Egres TR-2011-11}, 2012.

\bibitem{li2020deeprobust}
Yaxin Li, Wei Jin, Han Xu, and Jiliang Tang.
\newblock Deeprobust: A pytorch library for adversarial attacks and defenses.
\newblock {\em arXiv preprint arXiv:2005.06149}, 2020.

\bibitem{McAndrew21}
Duncan McAndrew.
\newblock The structural analysis of criminal networks.
\newblock {\em The Social Psychology of Crime}, 2021.

\bibitem{motwani1994computing}
Rajeev Motwani and Madhu Sudan.
\newblock Computing roots of graphs is hard.
\newblock {\em Discrete Applied Mathematics}, 54(1):81--88, 1994.

\bibitem{MOTWANI199481}
Rajeev Motwani and Madhu Sudan.
\newblock Computing roots of graphs is hard.
\newblock {\em Discrete Applied Mathematics}, 54(1):81--88, 1994.
\newblock URL: \url{https://www.sciencedirect.com/science/article/pii/0166218X94000239}, \href {https://doi.org/10.1016/0166-218X(94)00023-9} {\path{doi:10.1016/0166-218X(94)00023-9}}.

\bibitem{Muzio20}
Giulia Muzio, Leslie O’Bray, and Karsten Borgwardt.
\newblock Biological network analysis with deep learning.
\newblock {\em Briefings in bioinformatics}, 22(2):1515--1530, 2021.

\bibitem{pavlopoulos2011using}
Georgios~A Pavlopoulos, Maria Secrier, Charalampos~N Moschopoulos, Theodoros~G Soldatos, Sophia Kossida, Jan Aerts, Reinhard Schneider, and Pantelis~G Bagos.
\newblock Using graph theory to analyze biological networks.
\newblock {\em BioData mining}, 4:1--27, 2011.

\bibitem{networkrepo}
Ryan~A. Rossi and Nesreen~K. Ahmed.
\newblock The network data repository with interactive graph analytics and visualization.
\newblock In {\em AAAI}, 2015.
\newblock URL: \url{https://networkrepository.com}.

\bibitem{Nitai10}
Nitai~B. Silva, Ing-Ren Tsang, George D.~C. Cavalcanti, and Ing-Jyh Tsang.
\newblock A graph-based friend recommendation system using genetic algorithm.
\newblock In {\em IEEE Congress on Evolutionary Computation}, pages 1--7, 2010.

\bibitem{VANDAM2003241}
Edwin~R. {van Dam} and Willem~H. Haemers.
\newblock Which graphs are determined by their spectrum?
\newblock {\em Linear Algebra and its Applications}, 373:241--272, 2003.
\newblock Combinatorial Matrix Theory Conference (Pohang, 2002).
\newblock URL: \url{https://www.sciencedirect.com/science/article/pii/S002437950300483X}, \href {https://doi.org/10.1016/S0024-3795(03)00483-X} {\path{doi:10.1016/S0024-3795(03)00483-X}}.

\bibitem{van2023graph}
Piet Van~Mieghem.
\newblock {\em Graph spectra for complex networks}.
\newblock Cambridge university press, 2023.

\bibitem{VANMIEGHEM202434}
Piet {Van Mieghem} and Ivan Jokić.
\newblock Co-eigenvector graphs.
\newblock {\em Linear Algebra and its Applications}, 689:34--59, 2024.
\newblock URL: \url{https://www.sciencedirect.com/science/article/pii/S0024379524000521}, \href {https://doi.org/10.1016/j.laa.2024.02.008} {\path{doi:10.1016/j.laa.2024.02.008}}.

\bibitem{vilar2013detection}
Santiago Vilar, Eugenio Uriarte, Lourdes Santana, Nicholas~P Tatonetti, and Carol Friedman.
\newblock Detection of drug-drug interactions by modeling interaction profile fingerprints.
\newblock {\em PloS one}, 8(3):e58321, 2013.

\bibitem{Wilson09}
Christo Wilson, Bryce Boe, Alessandra Sala, Krishna~P.N. Puttaswamy, and Ben~Y. Zhao.
\newblock User interactions in social networks and their implications.
\newblock In {\em Proceedings of the 4th ACM European Conference on Computer Systems}, EuroSys '09, page 205–218, New York, NY, USA, 2009. Association for Computing Machinery.
\newblock \href {https://doi.org/10.1145/1519065.1519089} {\path{doi:10.1145/1519065.1519089}}.

\bibitem{Wu19}
Huijun Wu, Chen Wang, Yuriy Tyshetskiy, Andrew Docherty, Kai Lu, and Liming Zhu.
\newblock Adversarial examples for graph data: Deep insights into attack and defense.
\newblock In {\em Proceedings of the Twenty-Eighth International Joint Conference on Artificial Intelligence, {IJCAI-19}}, pages 4816--4823. International Joint Conferences on Artificial Intelligence Organization, 7 2019.
\newblock \href {https://doi.org/10.24963/ijcai.2019/669} {\path{doi:10.24963/ijcai.2019/669}}.

\bibitem{Xu23}
Xiaojun Xu, Hanzhang Wang, Alok Lal, Carl~A. Gunter, and Bo~Li.
\newblock Edog: Adversarial edge detection for graph neural networks.
\newblock In {\em 2023 IEEE Conference on Secure and Trustworthy Machine Learning (SaTML)}, pages 291--305, 2023.
\newblock \href {https://doi.org/10.1109/SaTML54575.2023.00027} {\path{doi:10.1109/SaTML54575.2023.00027}}.

\bibitem{zhang2017predicting}
Wen Zhang, Yanlin Chen, Feng Liu, Fei Luo, Gang Tian, and Xiaohong Li.
\newblock Predicting potential drug-drug interactions by integrating chemical, biological, phenotypic and network data.
\newblock {\em BMC bioinformatics}, 18:1--12, 2017.

\bibitem{Zugner18}
Daniel Z\"{u}gner, Amir Akbarnejad, and Stephan G\"{u}nnemann.
\newblock Adversarial attacks on neural networks for graph data.
\newblock In {\em Proceedings of the 24th ACM SIGKDD International Conference on Knowledge Discovery \& Data Mining}, KDD '18, page 2847–2856, New York, NY, USA, 2018. Association for Computing Machinery.
\newblock \href {https://doi.org/10.1145/3219819.3220078} {\path{doi:10.1145/3219819.3220078}}.

\end{thebibliography}

\appendix

\end{document}